# SECURE WEB-BASED STUDENT INFORMATION MANAGEMENT SYSTEM

By

**FALEBITA OLUWATOSIN SAMUEL**

**(NPA/03/02/00404)**

**A PROJECT SUBMITTED TO THE DEPARTMENT OF COMPUTER SCIENCE, FACULTY OF SCIENCE IN PARTIAL FULFILMENT OF REQUIREMENTS FOR THE AWARD OF BACHELOR OF SCIENCE (B.SC.) DEGREE IN COMPUTER SCIENCE NIGERIA POLICE ACADEMY, WUDIL, KANO.**

**OCTOBER, 2018**

# CERTIFICATION

This is to certify that this project has been read and approved as meeting the requirements of the Department of Computer Science and Mathematics, Nigeria Police Academy, Wudil, Kano, Nigeria for the Award of Bachelor of Science degree in Computer Science.

10/05/19

**Dr. Edward Philemon** **Date**

**(Supervisor)**

10-05-19

**Dr. Olorunpomi O.T.** **Date**

**(Head of Department)**

15/05/19

**Prof. A. A. Obiniyi** **Date**

**(External Examiner)**

# DEDICATION

This work is dedicated to my late grandfather Mr. Williams Falebita, for his last words of encouragement to me, may your beloved soul never cease to rest in perfect peace. Amen

# ACKNOWLEDGEMENT

Persian poet Saadi instructed “Have patience, all things are difficult before they become easy”; All glory and honor to the only wise God, my shalom, my sabaoth, my provider, my everything for the strength and grace he bestowed on me.

Never will I fail to express my profound appreciation to my project supervisor Mr. Edward Philemon for his valuable and constructive contribution during the entire process of my undergraduate degree and completion of my project. His willingness to give his time has been of a great motivation to me, also I would like to express my gratitude to Dr. Olorunpomi O.T. the Head, Computer Science Department, and the efforts of my unflagging lecturers in computer science are highly appreciated.

The success of everything is measure by the nature of its output, my family has been the fuel behind my engine, even when no one believed in me I still remember the sweet warm voice of Mr. and Mrs. Falebita, thanks for parenting me in this $21^{st}$ century in love with the fear of the lord, I am glad my achievement as given you a reason to smile and to my lovely siblings Falebita Omobolanle, Falebita Olamide, and Falebita Oladotun thanks for your support and prayers.

Rome wasn’t built in a day and I am pleased to express my gratitude to Professor Adefi Matthew Olong for his colossal contribution in building up my empire, to my sui generis ally Olayinka Ayanlade for his support, and to my significant other Ogundimbola Kenny Rachel thanks for all that you do.

# ABSTRACT

The reliability and success of any organization such as academic institution rely on its ability to provide secure, accurate and timely data about its operations, i.e. managing staff and students' information. Erstwhile managing student information in academic institution was done through paper-based information system, where academic records are documented in several files that are kept in shelves. Several problems are associated with paper-based information system. Managing information through the manual approach require physical exertion to retrieve, alter, and re-file the paper records. All this are nonvalue added services results in data inconsistency and redundancy. Currently several institutions have migrated to the automated web-based student information management system without considering the security architecture of the web portal. This project seeks to ameliorates and secure how information is being managed in Nigeria Police Academy through the development of a secured web-based student information management system, which has a friendly user interface that provides an easy and secure way to manage academic information such as students' information, staff information, course registration, course materials and results. This project was developed using Laravel 5.5 PHP Framework to provide a robust secure web-based student information system that is not vulnerable to 2018 OWASP TOP 10 web vulnerabilities.

# Table of Contents

# LIST OF TABLES

# LIST OF FIGURES

# CHAPTER ONE

## 1.1 Introduction

This chapter discuss the critical study of the existing system stating how information have been kept overtime, juxtaposed with the flaws in the system and provide a solution through the development of a secure web-based student information management system.

## 1.2 Background of the study

The Nigeria Police Academy relied heavily on paper-based information system for managing student's information usually carried out manually and kept in several files and shelves. This method of data management is obsolete and has many drawbacks. Firstly, paper-based information system is difficult to manage and track. The time required to retrieve, alter, and re-file the paper records are all non-value added activities. Moreover, keeping paper records consumes physical space, which leads to data inconsistency (different records of the same student appearing in different department/unitin the Academy at the same time), and data redundancy (having the same records in differentlocations). Secondly, information is disseminated to students via notice boards, and such information would require more time frame and to reach the intended students (Okibo & Ochiche, 2014).

Consequently, this project work seeks to design and implement a web-based student information management system that will provide an efficient alternative to the current paper-based approach to record keeping. The system utilizes Laravel PHP Framework that provides various mechanisms to secure a website such as: Encryption, Storing Password, Authenticating Users, Cross-site request forgery (CSRF), Avoiding SQL injection,

Protecting Routes, HTTP Basic Authentication which makes it a robust framework for secure information system development (Ali, 2018).

## 1.3 Statement of the Problem

The web-based student information management system can be deployed to maintain the records of students and staff easily. Achieving this objective is difficult using the paper-based information system approach as presently obtained in the academy, also this approachmakes it tedious to manage and track desired records. Moreover, it suffers from data redundancy and other anomalies, which makes it costlier to maintain and unreliable.

This project aim to design and implement a secure web-based student information management system that is secure and robust, which provides an efficient alternative to the current paper-based information system.

## 1.4 Aim and Objectives

The aim of this project is to develop a web-based student information management system that is secure and robust, which provides an efficient alternative to the current paper-based information system.

The specific Objectives are to:

i. Design and implement a web-based student information management system.

ii. Secure the web application using Laravel PHP Framework.

iii. Create a Student Portal using Laravel PHP Framework.

iv. Design an interface for lecturers to upload results and course materials.

## 1.5 Project Methodology

In other to achieve our stated objectives:

i. Related literature will be review in the domain of web application and information security.

ii. Develop a Web Application using the Full-Stack Web Development paradigm which is sub-divided into

  a. Front-end: The Front-end Framework utilized include Laravel Blade Template Engine, HTML 5, Bootstrap Framework v4.0.0-alpha.6 built on Cascading Style Sheet (CSS), Vue Js Framework built on JavaScript and npm package manager for installation of other Front-end libraries.

  b. Back-end: The Back-End Framework utilized include Laravel 5.5 Framework built on PHP v7.1.7.

  c. Database: Laravel Database backends that support using either raw SQL, the fluent query builder, and the Eloquent ORM (Object Relational Mapping) approach. Currently, Laravel supports four databases:

    i. MySQL.

    ii. PostgreSQL.

    iii. SQLite.

    iv. SQL Server.

iii. MySQL Database was selected for the database management of this system. The Software Development Methodology adopted in this project is the Object oriented analysis and design (OOAD). OOAD approach modules a system as a group of interacting objects. This methodology involves two stages; Object Oriented Analysis, and Object Oriented Design. Unified Modelling language (UML) notation is the design tool used for modelling in this project. The UML used in this project includes: Use case diagram, Activity diagram and Class Diagram.

iv. The implementation of this project was tested on this system configuration

a. Operating System: Windows 10 Pro.
b. RAM: 4GB.
c. Processor: Intel(R) Pentium(R) CPU P6200 @ 2.13GHz
d. System type: 64 bit Operating System, x64 based processor.

v. Other Tools utilized in this Web application development includes:

a. Laragon: For better performance optimization, “Laragon is the best and fastest local server by far” SniffleValve. Laragon v.3.1.6 localhost webserver was utilized for this project which consist of Apache/2.4.27 (Win64) OpenSSL/1.0.2l /PHP version: 7.1.7 and phpMyAdmin running MariaDB was used for the SQL database management Interface. (Valve, 2018)
b. Visual Studio Code: This is a lightweight but powerful source code editor which runs on Windows, macOsx and Linux. It comes with built-in support for JavaScript, TypeScript and Node.js and has a rich ecosystem of extensions for other languages (such as C++, C#, Java,

Python, PHP, Go) and runtimes (such as .NET and Unity). (Studio, 2018).

## 1.6 Significances of the study

Over the years, the academy has been bedeviled with the problem of maintaining students record, and access to relevant student information is usually cumbersome. This proposed system will provide a reliable and efficient way for managing student and staff details. Here are some significant achievements the proposed system is expected to provide. The system can assist the students have quick access to all courses that is to be taken throughout his/her undergraduate program, enable a faster access to their academic result and download of relevant course materials. Reduce the time needed to access any information about a student or staff, render faster and more convenient services to students, academic staff and non-academic staff. All the students' details will be process and send to a secure database, eliminating the previous paper-based information system.

## 1.7 Scope and limitations of the study

This is a non-generic web application developed for the Nigeria Police Academy, Wudil Kano (POLAC) suitable for managing Student and Staff Information. This system includes all faculties with their respective departments and keep track of all information about academic staff, non-academic staff and students. It can handle a series of task performed by the academic and non-academic staff hereby reducing staff workload. The system keep track of all students and staff records. It includes a faster means of record lookup which provides asynchronous query functionality using AJAX.

The system was unable to achieve the following

i. Provide an API that can be used by other developers to communicate with the website i.e. using Ionic-cordova to build an Andriod/IOS APP that will be compatible with the system ii. Automate the paper-based information system used in the Police wing of the academy.

iii. Provide a chatroom functionality for real time communication among users.

## 1.8 Project Outline

This project is structured into Five (5) chapters,

Chapter two discuss relevant literatures in domain of Web application and Information security and existing systems would also be considered alongside with their flaws.

Chapter three depicts the project methodology which entails all the stages of Software Development Life Cycle (SDLC). Also the interaction between the different component depict in several Unified Modeling Language (UML) such as the class diagram, activity diagram.

Chapter four discuss the implementation and design of the project using laravel framework to design and secure the application.

Chapter five gives the summary, recommendation and conclusion.

## 1.9 Definition of terms

Information Security (InfoSec): This is a set of strategies for managing the processes, tools and policies necessary to prevent, detect, document and counter threats to digital and non-digital information. (Security, 2016)

Paper-based Information System: This is used to describe a system that keeps information on paper rather than on a computer (Cambridge, 2019).

Laravel: This a free, open-source PHP web framework, created by Taylor Otwell and intended for the development of web applications following the model-view-controller (MVC) architectural pattern based on Symphony. (Otwell, 2019)

# CHATER TWO

# LITERATURE REVIEW

## 2.1 Introduction

This chapter discuss the history and state of the art of the web, web applications and web application vulnerabilities, also relevant literatures were review under the domain of information system security and web application security.

## 2.2 Web

According to (Berners-Lee, 2004), the World-Wide Web is a tangle of information that, through the implementation of hyperlinks, allows a browser to navigate usually quite randomly from one website to another. The meaning, context and applicability of the content of each Web page needs to be interpreted by the human reader.

Report from (Murdock, 2018) shows the current version of the web is Web 2.0, Tim Berners-Lee, director of World Wide Consortium and inventor of the World Wide Web in 1989. His notion of the “read-write” web is often used to describe Web 2.0 which as the ability to contribute content and interact with other web users.

Advancement in technology will someday leads us to Web 3.0 or Semantic Web. Berners-Lee’s goal is for the web to have agents, computer programs that have been written to collect web content from sources and communicate with other programs, in order to deliver the requested information to the user (Tim, Hendler, & Lassila, 2001).

### 2.2.1 Web Application

A Web application is a computer program that uses web browsers and web technology to perform a variety of operations over the internet. Web application use a combination of server-side scripts (PHP, Python and ASP) to handle the storage and retrieval of the information, and client-side scripts (JavaScript and HTML) to present information to users. This allows users to interact with the company using online forms, content management systems, shopping carts and more. In addition, the applications allow employees to create documents, share information, collaborate on projects and work on common documents regardless of communication (Henzel, 2018). Figure 2.1 depicts how a client access the web.

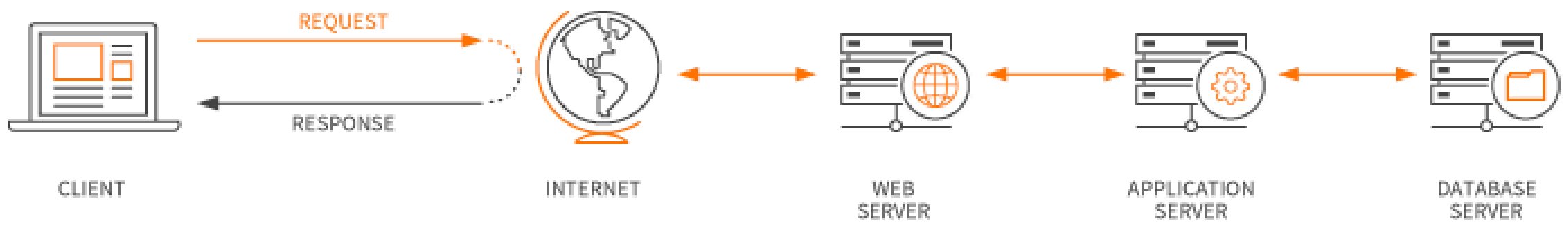

Figure 2. 1: Web application diagram

### 2.2.2 How Web Application Works

Web applications are usually coded in browser-supported language such as JavaScript and HTML as these languages rely on the browser to render the program executable. Some of the applications are dynamic, requiring server-side processing. Others are completely static with no processing required at the server. The web application requires a web server to manage requests from the client, an application server to perform the tasks requested, and, sometimes, a database to store the information. Application server technology ranges from ASP.NET, PHP and ColdFusion, to Python and JSP.

Here’s what a typical web application flow looks like:

1. User triggers a request to the web server over the Internet, either through a web browser or the application’s user interface.

2. Web server forwards this request to the appropriate web application server.
3. Web application server performs the requested task – such as querying the database or processing the data – then generates the results of the requested data.
4. Web application server sends results to the web server with the requested information or processed data.
5. Web server responds back to the client with the requested information that then appears on the user's display. (Henzel, 2018)

### 2.2.3 Web Application Vulnerabilities

The rapid growth in web application deployment has created more complex, distributed IT infrastructures that are harder to secure. For more than a decade, organizations have been dependent upon security measures at the perimeter of the network, such as firewalls, in order to protect IT infrastructures. However, now that more and more attacks are targeting security flaws in the design of web applications, such as injection flaws, traditional network security protection may not be sufficient to safeguard applications from such threats.

Web vulnerability is a weakness or misconfiguration in a website or web application code that enables an attacker (Hacker) to gain some level of control of the site and possibly the hosting server.

According to Open Web Application Security Project OWASP (Smithine, Stock, Gigler, & Glas, 2017) top 10 Application security risks as of 2017 are listed below.

i. Injection: Injection flaws, such as SQL, OS, and LDAP injection occur when untrusted data is sent to an interpreter as part of a command or query. The attacker’s hostile data can trick the interpreter into executing unintended commands or accessing data without proper authorization.

ii. Broken Authentication: Application functions related to authentication and session management are often implemented incorrectly, allowing attackers to compromise passwords, keys, or session tokens, or to exploit other implementation flaws to assume other users’ identities (temporarily or permanently).

iii. Sensitive Data Exposure: Many web applications and APIs do not properly protect sensitive data, such as financial, healthcare, and PII. Attackers may steal or modify such weakly protected data to conduct credit card fraud, identity theft, or other crimes. Sensitive data deserves extra protection such as encryption at rest or in transit, as well as special precautions when exchanged with the browser.

iv. XML External Entity (XXE): Many older or poorly configured XML processors evaluate external entity references within XML documents. External entities can be used to disclose internal files using the file URI handler, internal SMB file shares on unpatched Windows servers, internal port scanning, remote code execution, and denial of service attacks, such as the Billion Laughs attack.

v. Broken Access Control: Restrictions on what authenticated users are allowed to do are not properly enforced. Attackers can exploit these flaws to access unauthorized functionality and/or data, such as access other users' accounts, view sensitive files, modify other users’ data, change access rights, etc. vi. Security Misconfiguration: Security misconfiguration is the most common issue in the data, which is due in part to manual or ad hoc configuration (or not configuring at all), insecure default configurations, open S3 buckets, misconfigured HTTP headers, error messages containing sensitive information, not patching or upgrading systems, frameworks, dependencies, and components in a timely fashion (or at all).

vii. Cross-Site Scripting (XSS): XSS flaws occur whenever an application includes untrusted data in a new web page without proper validation or escaping, or updates an existing web page with user supplied data using a browser API that can create JavaScript. XSS allows attackers to execute scripts in the victim's browser which can hijack user sessions, deface web sites, or redirect the user to malicious sites.

viii. Insecure Deserialization: Insecure deserialization flaws occur when an application receives hostile serialized objects. Insecure deserialization leads to remote code execution. Even if deserialization flaws do not result in remote code execution, serialized objects can be replayed, tampered or deleted to spoof users, conduct injection attacks, and elevate privileges.

ix. Using Components with Known Vulnerabilities: Components, such as libraries, frameworks, and other software modules, run with the same privileges as the application. If a vulnerable component is exploited, such an attack can facilitate serious data loss or server takeover. Applications and APIs using components with known vulnerabilities may undermine application defenses and enable various attacks and impacts.

x. Insufficient Logging & Monitoring: Insufficient logging and monitoring, coupled with missing or ineffective integration with incident response allows attackers to further attack systems, maintain persistence, pivot to more systems, and tamper, extract or destroy data. Most breach studies show time to detect a breach is over 200 days, typically detected by external parties rather than internal processes or monitoring.

### 2.2.4 Web Application Security

Web application security has been a major threat in information technology since the evolvement of dynamic web application. (Yadav, 2014) suggested that some threats originate from non-trusted client access points, session-less protocols, the general complexity of web technologies, and network-layer insecurity. With web applications, client software usually cannot always be controlled by the application owner. Therefore, input from a client running the software cannot be completely trusted and processed directly. An attacker can forge an identity to look like a legitimate client, duplicate a user’s identity, or create fraudulent messages and cookies. In addition, HTTP is a session-less protocol, and is therefore susceptible to replay and injection attacks. Hypertext Transport Protocol messages can easily be modified, spoofed and sniffed.

## 2.3 Information System

An Information System (IS) can be any organized combination of people, hardware, software, communications networks, data resources, policies and procedures that stores, retrieves, transforms and disseminates information in an organization. (O'Brien & Marakas, 2011).

### 2.3.1 Information System Security

There is a close link between Information and Security and it is clearly established by the fact that the information of the company is as reliable as the strength of the security system designed to protect the information. If the security system is not effective in protecting the information, then there would be a sense of mistrust and uncertainty about the information emerging from that system and that would definitely not have a positive impact on the business. On the opposite if the company has a strong security system the information is termed reliable and it would benefit business from both outside and inside. (Alghazzawi , Hasan , & Trigui, 2014) .

One of the most important asset of an organization in today's world of increasing dependence on technology and the application of IT in almost all the spheres of business, is Information. It is impertinent that an organization manages its information with utmost care and diligence. The criticality of information can be compared with that of work or capital and at times even more as with the advent of technology modern startups are completely based on information and it is the core product of the business. In reality, the number of organizations getting dependent greatly on IS (Information System) is ever increasing over the past few years. (Mellado, Fernandez-Medina, & Piattini, 2007). The role of Information Systems in the world today is widely being accepted and they are at the center of almost all the technology infrastructures related to critical functions and the same is recognized by the researchers in the field of security and technology. (Mellado, Blanco, Sanchez, & Fernandez-Medina, 2010).

### 2.3.2 Information System Vulnerabilities

We are aware that the online based information management are the targets of cyberattacks from a varietyof malicious cyber actors ranging from hackers, to cyber terrorists, to viruses on the internet, to insider threat fromemployees of the company or even phishing through socially engineered attacks. (Choo, Smith, & McCusker, 2007).

The requirements of security in technology have been on the rise ever since the 70s century and this has led to the development of a vast Security Protocols, Models and Techniques. Development of the security tools has also made the international communitypay attention to developing of international certifications standards. In fact, it is so noticeable, as highlighted in (ITU, 2019) ICT Security Standards Roadmap International Telecommunication Union that we can today find a number of international organizations that have laid down complex arrangement of standards and benchmarks related to the field of information security and even these standards are constantly updated & changed as required.

However, with the serious threat of unauthorized users on the internet, Information System Security (ISS) is facing unprecedented challenges and effective Information

System Security Management (ISSM) is one of the major concerns (Hone & Eloff, 2002). Criminals, terrorists, disgruntled employees, technical problems and many other issues can threaten the security and integrity of information systems (Nissenbaum, 2005). Given the importance of information stored in these systems, it is reasonable to believe that information systems security should be an important managerial concern, as much of the literature suggests (Siponen, 2005). ISS is perceived as a way of fighting and preventing criminal activities (EC, 2007). Hacking, malware and viruses constitute problems that security needs to check (Turner & Broucek, 2003). However, this connects ISS with law enforcement and in particular with digital forensics (Sitaraman & Venkatesan, 2006). There are many challenges in maintaining security in higher learning institutions (Doherty & Fulford, 2006) which deteriorates the use of information system in universities.

## 2.4 Application of Information System in Universities

In view of (Okibo & Ochiche, 2014), Information security challenge in higher education is limited budgets especially in today's economy. Another occurring challenge is the cultural adaptation to academic information security management. Higher education environments typically have several departments that utilize information technology in separate facilities; from faculty to students; deans to VCs of academic affairs; each has the challenge with balancing information security and an end-user happiness. It's practically impossible given all the pressure.

Universities are relying in information systems to carry out their day to day operations. More specifically is the use of Academic Management Systems (AMS) by numerous universities for their business operations including teaching, student administration, research and development. Information security application to university's ISs is strategically important to maintaining overall business continuity. The ever emerging threats that are experienced with preservation of information through databases are made more exquisite and different with each threat being as complicated as one can think of securing (Adamkiewicz, 2005). To effectively manage information in a higher learning institution's context involves the process of applying information security to ensure risks, finances and efforts are balanced while at the same time continuous learning and improvements are cultured (Gefen, 2004). Security should be the concern of everyone in the organization and it should be a way of life within the institution's fraternity.

Universities have adopted information systems and the related technologies so as to gain a competitive edge. In this era effective control of operations and strong strategies are associated with management of quality information. The aspect of readily available information means that universities are affected by their dependence on information and technology resources, systems and the underlying structures that form the basis for this technologies and systems. In universities, reliance on information systems is evident on activities related to creating, using and sharing of information in teaching, learning, research and development and when marketing the university through its websites. It's evident that the amount of intellectual property generated by universities and importance of university information is extensive. The demand for effective information security management is ultimately a combination of various related factors. These factors comprise of reliance of information, increase in the threats that hinder the information that is relied upon heavily and the need for the controls to reduce this ever emerging new risks. Currently there is limited published academic literature that emphasizes on information security management in higher learning. Most of the literature analyzed so far focuses on information security management in organizations and not universities.

## 2.5 Securing Information System

It is expedient to secure our student information management system to improve the security of web applications, an open and freely-accessible community called the Open Web Application Security Project (OWASP) has been established to coordinate worldwide efforts aimed at reducing the risks associated with web application software.

The major area of this project is to develop a Secure Web-based Student Information Management System that can resist some web application vulnerabilities for managing Student Information in Nigeria Police Academy Wudil, Kano. The fastest growing PHP Framework in 2018 called LARAVEL was used to developed the web application, as at the time of developing this project the stable version of Laravel is 5.5, all security functionality of Laravel were utilized to ensure the security of the Student Information Management System.

## 2.6 Review of related works

Several work have been done in the area of Student Information Management System (SIMS)

- (Gunathilake, Indrathilake, & Wedagedera, 2009) proposed an opensource webbased MIS for the University of Ruhuna, Sri Lanka. This they were able to implement with the LAMP/WAMP technologies. They were able to categorize their users based on administrator, super admin, top admin, general, lecturer and student. The pilot version was targeted at their Faculty of Science and they achieved a password encryption with the primary DES algorithm.

- (Mariusz , 2010) in his solution University Study-Oriented System (USOS) in Poland stated that the main functional parts are the admin, web, admission/registration of students, database of results, course and diploma catalog, statistics etc. According to him, this solution is used by 27 higher education Polish institutions. In such a system, before transferring any module for production use it has to pass through sample database and university test. Documentation comprising system specification and implementation were updated regularly. Such solutions enhance communication between students and lecturers, proper security measures to prevent against Cross-site request forgery.
- (Bharaagoudar, Geeta, & Totad, 2013) developed a web-based Student Information Management System in India which could send emails to students to validate their mailbox on registration. They were able to achieve this using technology such as HTML, CSS, Javascript, PHP and SQL. According to their description, it is a paperless work that assists in automating existing manual methods and can be remotely monitored and controlled on a server based network, the SIMS developed had no built-in security measures to prevent SQL injection.
- (Hemn & Wu , 2014) proposed a system in China that can provide students' general and educational information. According to them, the Students Information Management System (SIMS) can be used to create, read and update the details of a student and also generate reports about his/her skills and experience. Such systems save time of retrieval and prevent data loss.

In a publication by (Charletta, 2004) at North Illinois University USA, he noted a lawsuit filed against Microsoft by a lady in Los Angeles over security holes in the company's software. The plaintiff was a film maker, Marcy Hamilton who charged that because of shoddy workmanship by Microsoft, she had become a victim of identity theft. According to her, her Social Security Number (SSN) and bank information was stolen online. Hence, in this project, we have greatly considered issues of security breaches and have recently incorporated bycrpt hashing algorithm to secure all sensitive information and enable access control list (ACL) to manage the permission of every user, limiting the information available to each designate member. This will make the system more credible and enable the management to control information based on their assigned roles. Several information systems accessed through the internet, this information system is developed as web application that can be access remotely ranging from the use of personal computer (PC) to smart phones.

# CHAPTER THREE

# DESIGN OF STUDENT INFORMATION MANAGEMENT SYSTEM

## 3.1 Introduction

This chapter critically examines the project methodology which entails all the stages of Software Development Life Cycle (SDLC). Also the interaction between the different component is been depicted in several Unified Modeling Language (UML) tools such as the class diagram, activity diagram.

## 3.2 System Design

Methodology can be defined as consisting of phases which will guide systems developers in their choice of techniques at each stage of a project, to help in the planning, management, control and evaluation of the system or project. With respect to information systems, it is acollection of procedures, techniques, tools and documentation aid which will help the systems developers in their efforts to implement a new information system. Meanwhile, due to the nature of this project, the specific methodology used is this project is the Object-Oriented analysis and design (OOAD).

## 3.3 Architectural Design

The Software Development Methodology adopted in this project is the Object-oriented analysis and design (OOAD). OOAD approach modules a system as a group of interacting object. This methodology involves two stages; Object Oriented Analysis and Object Oriented Design. Object Modeling is somewhat similar to the traditional approach of system designing, in that it also follows sequential process of system designing but with different approach. The basic steps of system designing using Object Modeling may involve:

i. System Analysis.
ii. System Design.
iii. Object Design.
iv. Implementation.

Unified Modelling language (UML) notation is the design tool used for Object modeling in this project. The UML used in this project includes: use case diagrams, class diagrams, sequence diagrams, state transition diagrams, and activity diagrams.

## 3.4 System Analysis Phase

This is a phase with the purpose of analyzing current system. In other words, it is process of gathering and interpreting facts about the current system in order to understand it. Diagnosing its difficulties and using those facts to improve the system through better procedures and methods. In any system development, this phase plays a vital role in the realization of the system. The investigations are necessary as they provide the system design phase with an overview of the kind of data that would serve as the input. Output and the processing done on the data in the new system to be developed.

The systems analysis phase includes the four main activities shown in Figure 3.1: requirements modeling, data and process modeling, object modeling, and consideration of development strategies.

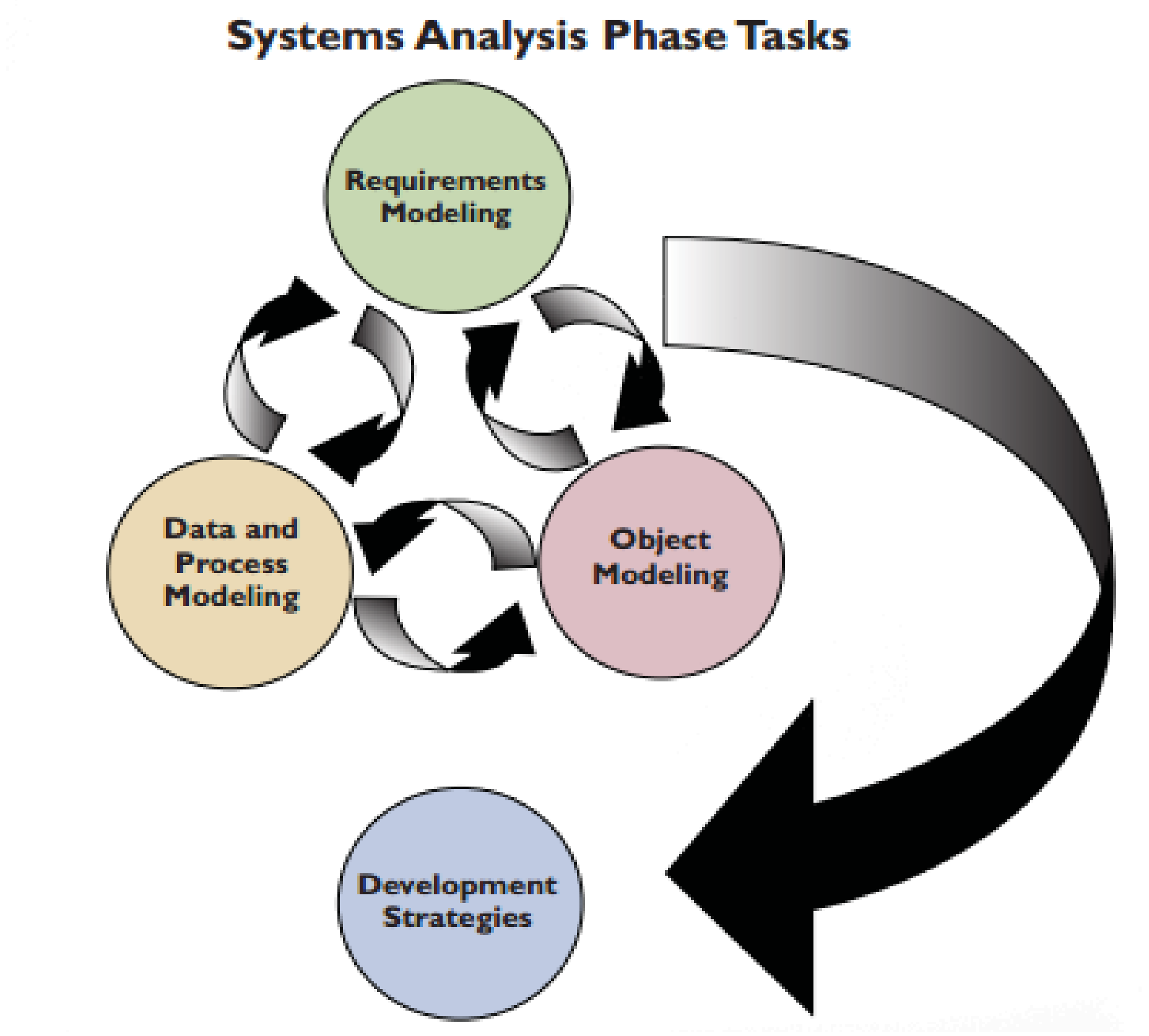

Figure 3. 1: The systems analysis phase

### 3.4.1 Requirements Modeling

This stage involves fact-finding to describe the current system and identification of the requirements for the new system, such as outputs, inputs, processes, performance, and security.

#### 3.4.1.1 Fact Finding About the Current System

A thorough investigation of the current system was carried out in order to obtain detailed information about the application area to be designed. In the course of my findings, several effective methods of information gathering or data collection were employed which include: interviewing the examination officers; discussion with pertinent stakeholders of the system such as the HODs and lecturers; evaluation and inspection of relevant documents such as course registration form, bio-data form, and staff bio-data form.

#### 3.4.1.2 Overview of The Current System

The system starts with registration of new staff and students manually using a paper based approach to obtain information. Student are provided with their required courses for each semester which is also done through a paper based approach. Courses are to be allocated to the lecturers by the Head of the Department. Lecturers enters corresponding subject's attendance and marks of a student will be entered in the Excel sheets and validations are to be done by the user itself. A number of risk is involved, moreover a lot of work needs to be done and the user must be conscious when entering the details into Ms. Excel.

#### 3.4.1.3 Problems in The Current System

Paper-based information system is difficult to manage and track. The physical exertion required to retrieve, alter, and re-file the paper records are all non-value added activities. Moreover, keeping paper records consumes physical space, which leads to data inconsistency (different records of the same student appearing in different department/ unit in the Academy at the same time), and data redundancy (having the same record in different locations). Whenever information is disseminated to students via notice boards, such information may take longer time to reach the intended recipients.

#### 3.4.1.4 Requirement for The New System

Requirement determination involves the study of the current system to find out how it works and where improvements should be made. A requirement therefore is any feature that must be included in the new system to improve the present situation that may include the way of capturing or processing data. Producing information and so on. The success of the new system however depends on the correctness of the requirement obtain. The combination of interview was used to provide information about the system.

### 3.4.2 Object Modeling

Object-oriented analysis (O-O) combines data and the processes that act on the data into things called objects. These objects represent actual people, things, transactions, and events that affect the system (Shelly & Rosenblatt, 2009). During the system development process, analysts often use both modeling methods to gain as much information as possible. The Objects in this project includes:

i. Staff.
ii. Cadet.
iii. Admin.
iv. Courses.

#### 3.4.2.1 Unified Modeling Language

UML is a widely used method of visualizing documenting and to develop object models of an information system. The objective of the Unified Modeling Language is to provide a common vocabulary of object-based terms and diagramming techniques that is rich enough to model any systems development project from analysis to design. The current version of UML, version 2.0, was accepted by the Object Management Group (OMG) in 2003. The UML used in this project includes: use case diagrams, class diagrams and activity diagrams.

#### 3.4.2.2 Use Case Diagram

A use case diagram illustrates in a very simple way the main functions of the system and the different kinds of users who will interact with it. The Use Case is the foundation of UML, and the Use Case Diagram contains the use cases.

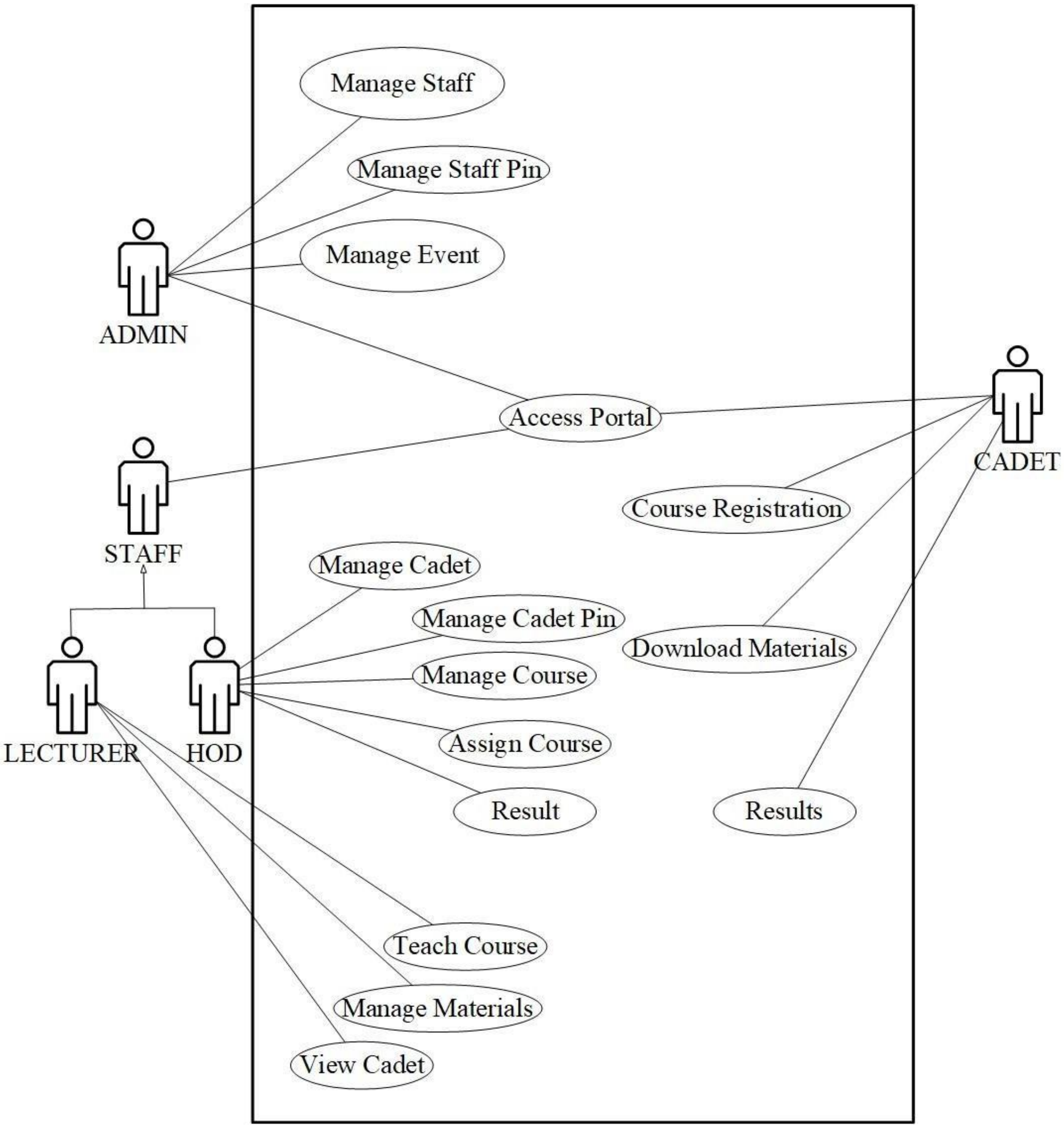

Figure 3. 2: Secure Student Information Management System Use Case Diagram

#### 3.4.2.3 Activity Diagram

Activity diagrams are used to model the behavior in a business process independent of objects. Activity diagrams can be used to model everything from a high-level business workflow that involves many different use cases, to the details of an individual use case, all the way down to the specific details of an individual method. In a nutshell, activity diagrams can be used to model any type of process.

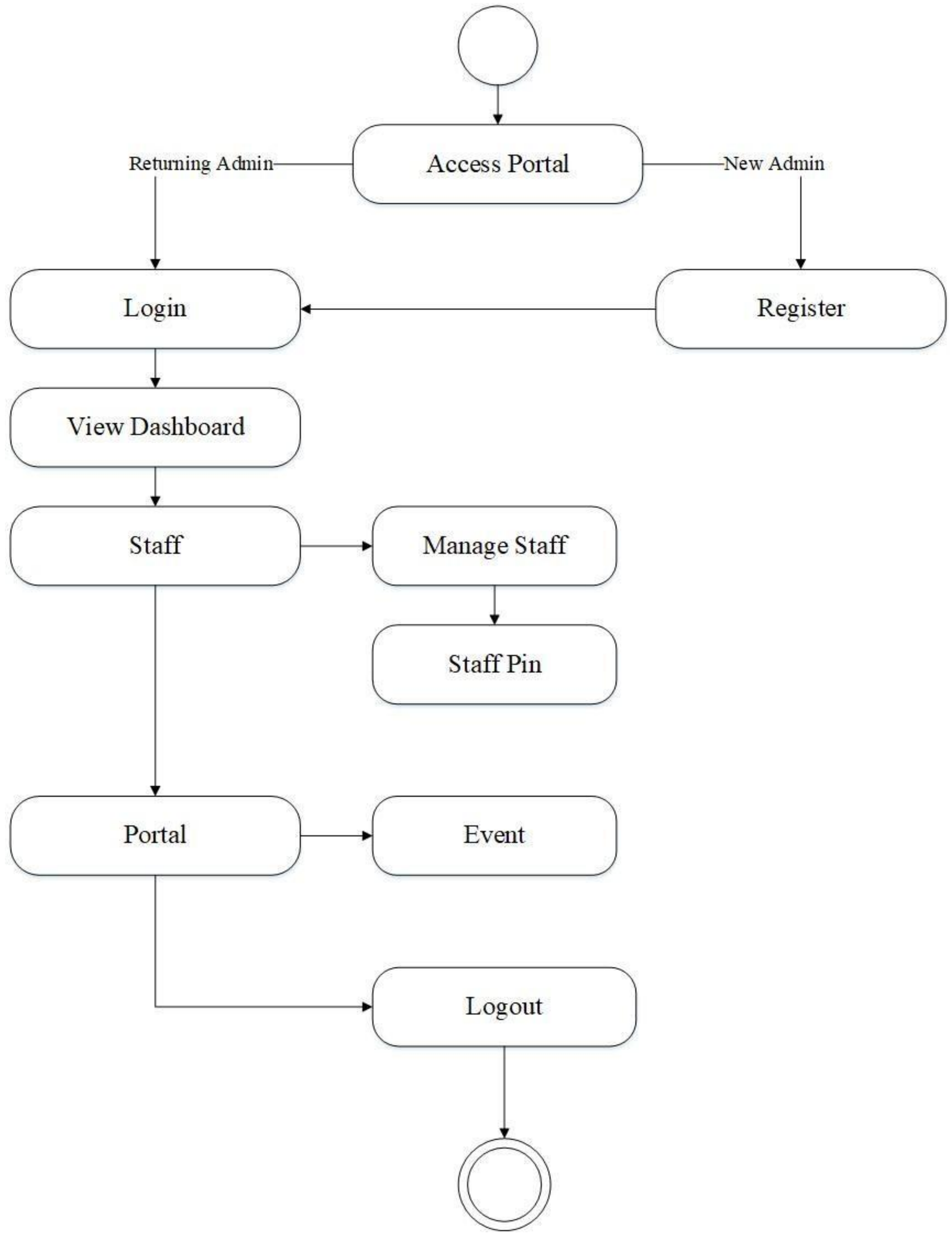

Figure 3. 3: Activity Diagram for Admin.

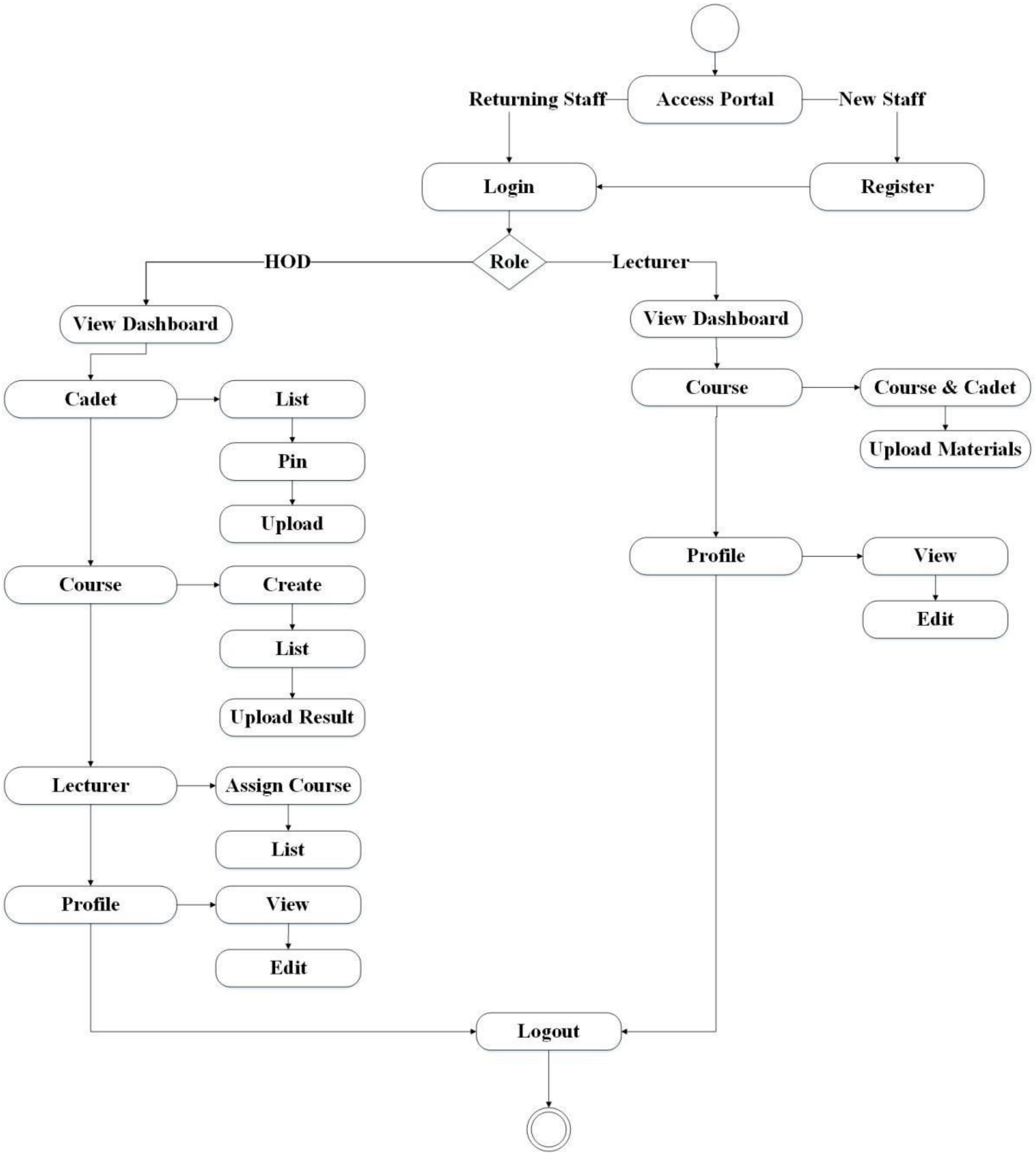

Figure 3. 4: Activity Diagram for Staff.

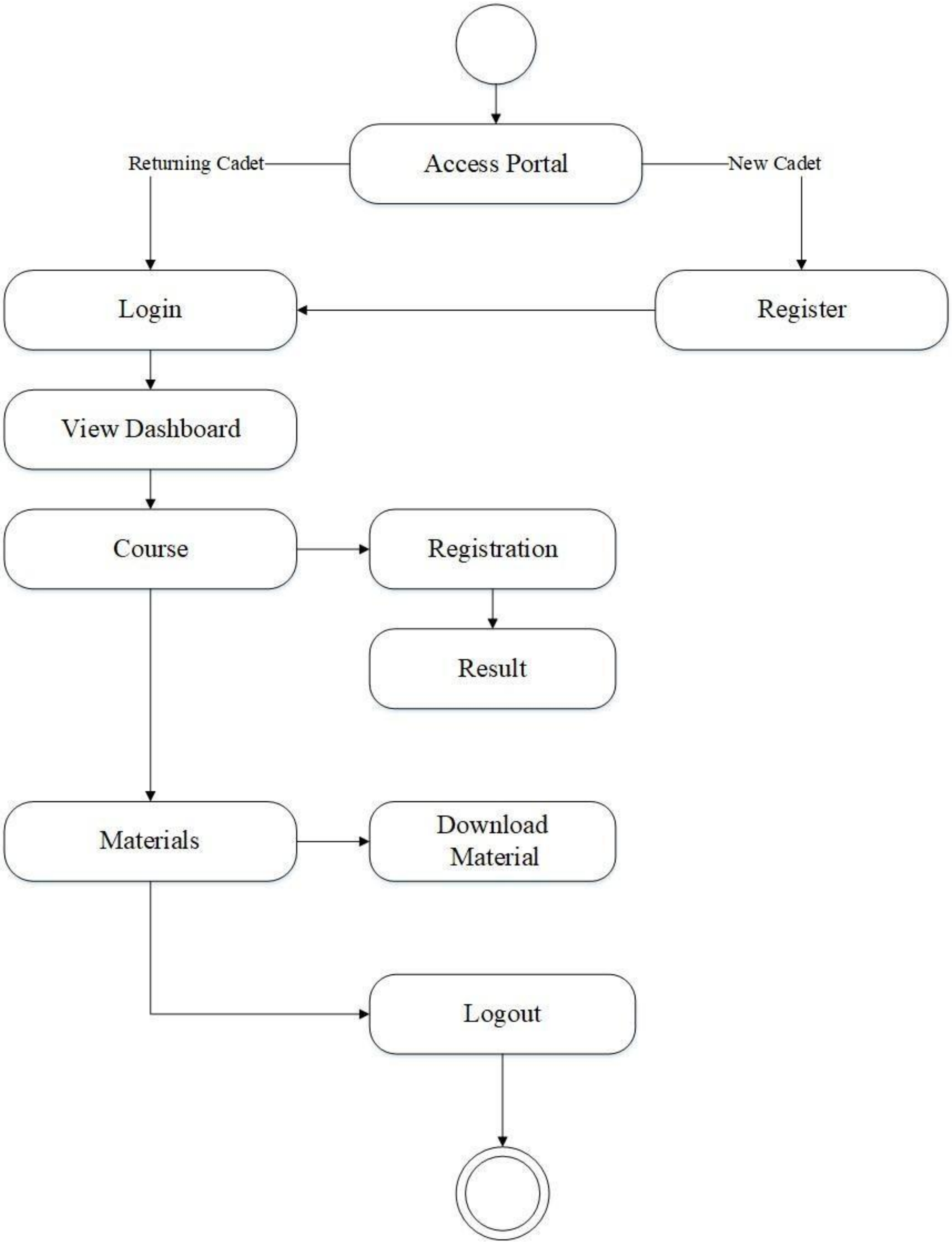

Figure 3. 5: Activity Diagram for Cadet

### 3.4.2.4 Class Diagram

Class diagram shows the object class in the system and association between these classes. The class diagram describes the types of objects in a system and the various kinds of static relationships that exist among them. In UML, a class is represented by a rectangle with one or more horizontal compartments. The upper compartment holds the name of the class. The name of the class is the only required field in a class diagram. By convention, the class name starts with a capital letter. The (optional) center compartment of the class rectangle holds the list of the class attributes/data members, and the (optional) lower compartment holds the list of operations/methods.

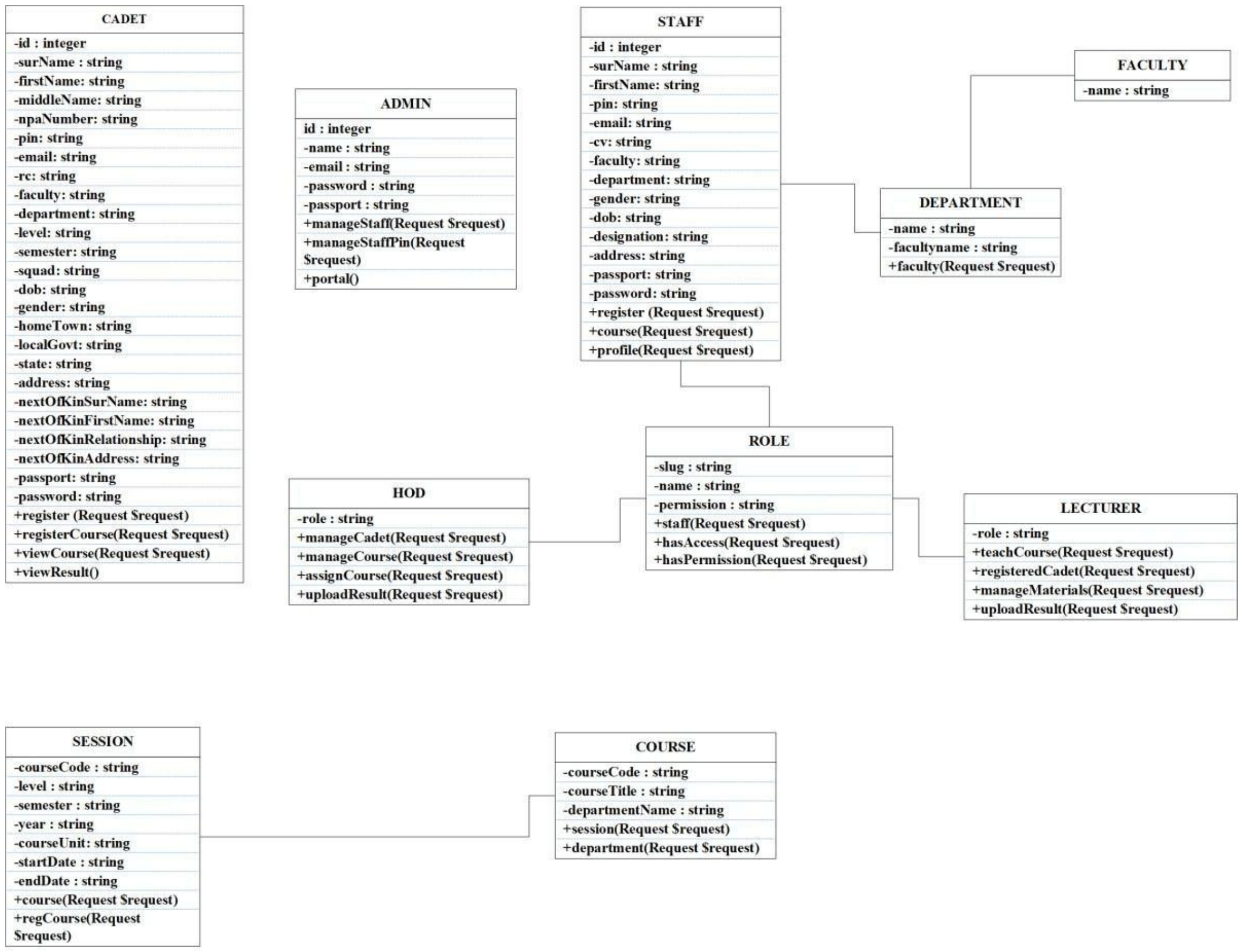

Figure 3. 6: Class diagram for Student Information Management System.

## 3.5 System Design Phase

The design phase refines the analysis model and applies the needed technology and other implementation constrains. The purpose of design is to decide how to build the system. Design also includes activities such as designing the user interface, system inputs, and system outputs, which involve the ways that the user interacts with the system.

### 3.5.1 Architecture Design

An important component of the design phase is the architecture design. These involves four major functions (data storage, data access logic, application logic, and presentation logic). The system design utilized a three-tiered architecture which include the use of three sets of computers, as shown in (Figure 3.7)

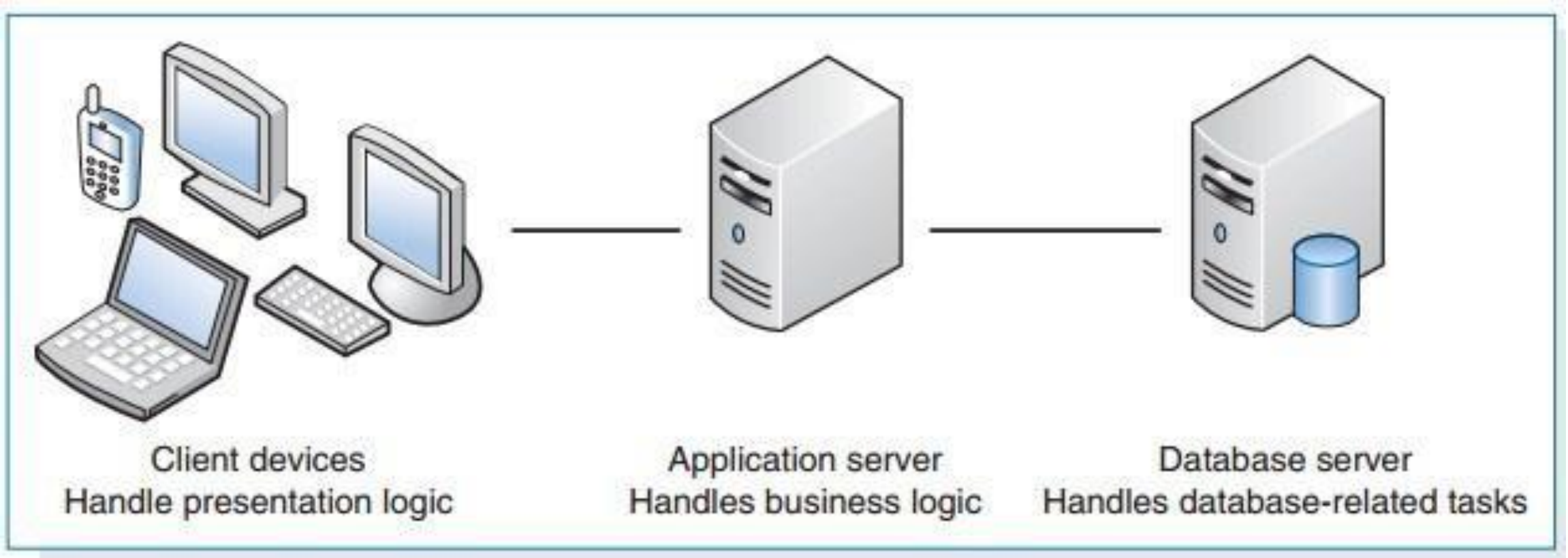

Figure 3. 7: Three-Tiered Client–Server Architecture.

In this case, the software on the client computer is responsible for presentation logic, an application server(s) is responsible for the application logic, and a separate database server(s) is responsible for the data access logic and data storage. Which follows the Full-Stack Web Development paradigm that is divided into 3 layers (Frontend, Backend and Database).

i. Frontend: This mostly focus on what some may coin the "Client side" of development and it handles the presentation logic. It's what the user sees, touches and experiences, the three main languages use for front-end development are Hyper Text Markup Language (HTML), Cascading Style Sheet (CSS) and JavaScript.

ii. Backend: This basically handles the business logic and how the site works, updates, and changes. This refers to everything the user can't see in the browser, like databases interaction and server. It's also known as "server side". The major languages use for backend development are PHP, Python, Java and Ruby.

iii. Database: A database is a collection of information that is organized so that it can be easily accessed, managed and updated. There are many different kinds of databases, ranging from the most prevalent approach, the relational database, to a distributed database, cloud database or NoSQL database. (Target, 2018)

### 3.5.2 Database Design

Database is critical for all businesses. A good database does not allow any form of anomalies and stores only relevant information in an ordered manner. If a database has anomalies, it is affecting the efficiency and data integrity. For example, delete anomaly arise upon the deletion of a row which also forces other useful data to be lost. As such, the tables need to be normalized. This fulfils the last objective of ensuring data are accurate and retrieved correctly. For the database of this project, the tables are normalized to BCNF as shown in (Table 3.1-3.8)

| # | Name | Type | Collation | Attributes | Null | Default |
|---|---|---|---|---|---|---|
| 1 | id | int(10) | | UNSIGNED | No | *None* |
| 2 | surName | varchar(191) | utf8mb4_unicode_ci | | No | *None* |
| 3 | firstName | varchar(191) | utf8mb4_unicode_ci | | No | *None* |
| 4 | faculty | varchar(191) | utf8mb4_unicode_ci | | No | *None* |
| 5 | department | varchar(191) | utf8mb4_unicode_ci | | No | *None* |
| 6 | pin | varchar(191) | utf8mb4_unicode_ci | | No | *None* |
| 7 | passport | varchar(191) | utf8mb4_unicode_ci | | No | avatar.png |
| 8 | cv | varchar(191) | utf8mb4_unicode_ci | | Yes | *NULL* |
| 9 | designation | varchar(191) | utf8mb4_unicode_ci | | Yes | *NULL* |
| 10 | address | varchar(191) | utf8mb4_unicode_ci | | Yes | *NULL* |
| 11 | email | varchar(191) | utf8mb4_unicode_ci | | No | *None* |
| 12 | dob | varchar(191) | utf8mb4_unicode_ci | | Yes | *NULL* |
| 13 | password | varchar(191) | utf8mb4_unicode_ci | | No | *None* |
| 14 | remember_token | varchar(100) | utf8mb4_unicode_ci | | Yes | *NULL* |
| 15 | created_at | timestamp | | | Yes | *NULL* |
| 16 | updated_at | timestamp | | | Yes | *NULL* |

Table 3. 1: Student Table

| Keyname | Type | Unique | Packed | Column | Cardinality | Collation | Null | Comment |
|---|---|---|---|---|---|---|---|---|
| PRIMARY | BTREE | Yes | No | id | 0 | A | No | |
| staff_email_unique | BTREE | Yes | No | email | 0 | A | No | |
| staff_department_index | BTREE | No | No | department | 0 | A | No | |

Table 3. 2: Student Table Index

| # | Name | Type | Collation | Attributes | Null | Default |
|---|---|---|---|---|---|---|
| 1 | id | int(10) | | UNSIGNED | No | *None* |
| 2 | surName | varchar(191) | utf8mb4_unicode_ci | | No | *None* |
| 3 | firstName | varchar(191) | utf8mb4_unicode_ci | | No | *None* |
| 4 | faculty | varchar(191) | utf8mb4_unicode_ci | | No | *None* |
| 5 | department | varchar(191) | utf8mb4_unicode_ci | | No | *None* |
| 6 | pin | varchar(191) | utf8mb4_unicode_ci | | No | *None* |
| 7 | passport | varchar(191) | utf8mb4_unicode_ci | | No | avatar.png |
| 8 | cv | varchar(191) | utf8mb4_unicode_ci | | Yes | *NULL* |
| 9 | designation | varchar(191) | utf8mb4_unicode_ci | | Yes | *NULL* |
| 10 | address | varchar(191) | utf8mb4_unicode_ci | | Yes | *NULL* |
| 11 | email | varchar(191) | utf8mb4_unicode_ci | | No | *None* |
| 12 | dob | varchar(191) | utf8mb4_unicode_ci | | Yes | *NULL* |
| 13 | password | varchar(191) | utf8mb4_unicode_ci | | No | *None* |
| 14 | remember_token | varchar(100) | utf8mb4_unicode_ci | | Yes | *NULL* |
| 15 | created_at | timestamp | | | Yes | *NULL* |
| 16 | updated_at | timestamp | | | Yes | *NULL* |

Table 3. 3: Staff Table

| Keyname | Type | Unique | Packed | Column | Cardinality | Collation | Null | Comment |
|---|---|---|---|---|---|---|---|---|
| PRIMARY | BTREE | Yes | No | id | 0 | A | No | |
| staff_email_unique | BTREE | Yes | No | email | 0 | A | No | |
| staff_department_index | BTREE | No | No | department | 0 | A | No | |

Table 3. 4: Staff Table Index

| # | Name | Type | Collation | Attributes | Null | Default |
|---|---|---|---|---|---|---|
| 1 | id | int(10) | | UNSIGNED | No | *None* |
| 2 | name | varchar(191) | utf8mb4_unicode_ci | | No | *None* |
| 3 | email | varchar(191) | utf8mb4_unicode_ci | | No | *None* |
| 4 | passport | varchar(191) | utf8mb4_unicode_ci | | No | avatar.png |
| 5 | password | varchar(191) | utf8mb4_unicode_ci | | No | *None* |
| 6 | remember_token | varchar(100) | utf8mb4_unicode_ci | | Yes | *NULL* |
| 7 | created_at | timestamp | | | Yes | *NULL* |
| 8 | updated_at | timestamp | | | Yes | *NULL* |

Table 3. 5: Admin Table

| Keyname | Type | Unique | Packed | Column | Cardinality | Collation | Null | Comment |
|---|---|---|---|---|---|---|---|---|
| PRIMARY | BTREE | Yes | No | id | 0 | A | No | |
| admins_email_unique | BTREE | Yes | No | email | 0 | A | No | |

Table 3. 6: Admin Table Index

| # | Name | Type | Collation | Attributes | Null | Default |
|---|---|---|---|---|---|---|
| 1 | course_code | varchar(191) | utf8mb4_unicode_ci | | No | *None* |
| 2 | course_title | varchar(191) | utf8mb4_unicode_ci | | No | *None* |
| 3 | dept_name | varchar(191) | utf8mb4_unicode_ci | | No | *None* |
| 4 | created_at | timestamp | | | Yes | *NULL* |
| 5 | updated_at | timestamp | | | Yes | *NULL* |
| 6 | deleted_at | timestamp | | | Yes | *NULL* |

Table 3. 7: Course Table

| Keyname | Type | Unique | Packed | Column | Cardinality | Collation | Null | Comment |
|---|---|---|---|---|---|---|---|---|
| PRIMARY | BTREE | Yes | No | course_code | 0 | A | No | |
| courses_dept_name_index | BTREE | No | No | dept_name | 0 | A | No | |

Table 3. 8: Course Table Index

#### 3.5.2.1 Database Schema

A database schema, along with primary key and foreign key dependencies, can be depicted by schema diagrams. Figure 3.8 shows the schema diagram

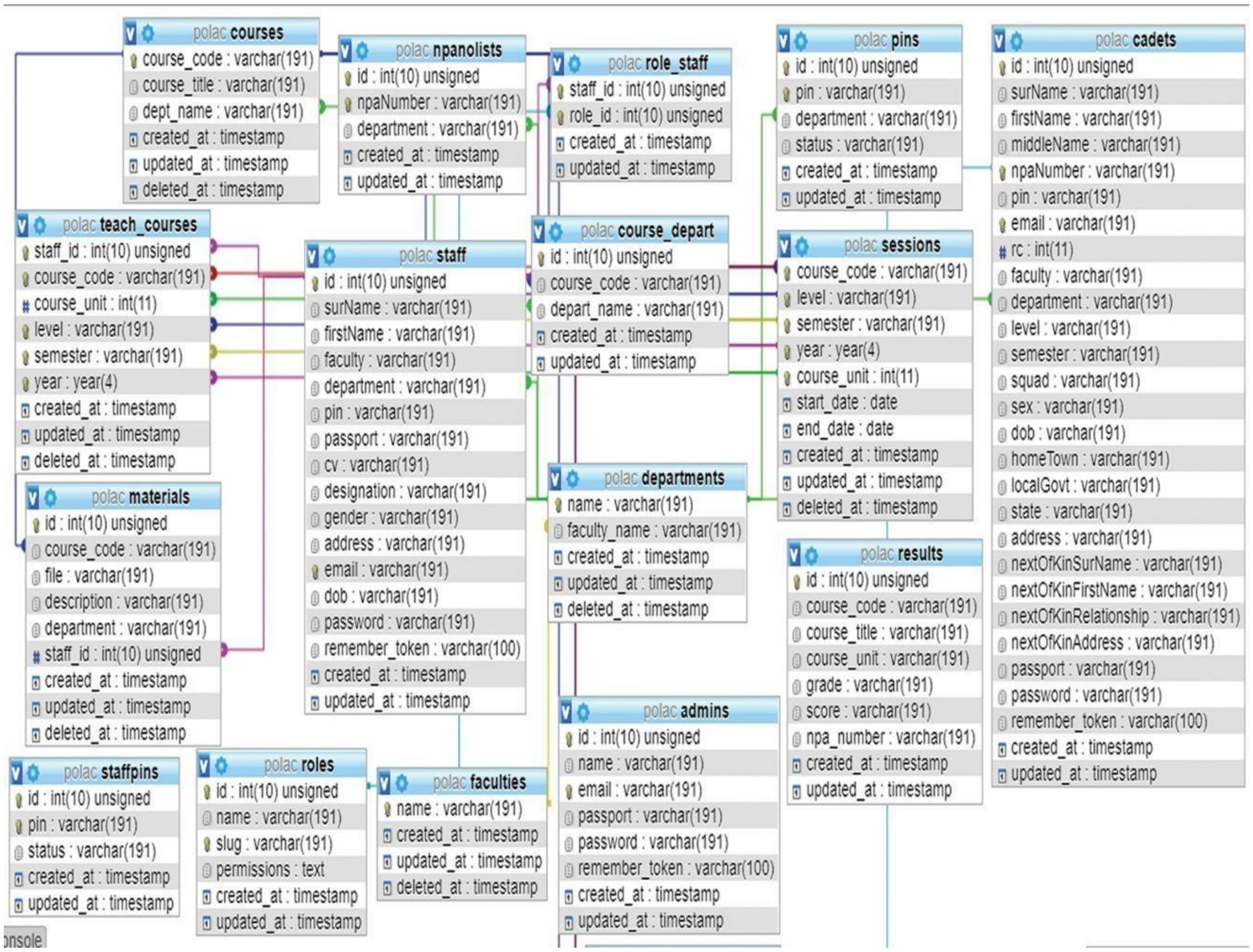

Figure 3. 8: Schema Diagram for Secure Student Information Management System.

### 3.5.2.2 Database E-R Diagram

An entity-relationship diagram (ERD) is a model that shows the logical relationships and interaction among system entities. An ERD provides an overall view of the system and a blueprint for creating the physical data structures. Figure 3.9 shows the entity-relationship diagram for the Secure Student Information Management System.

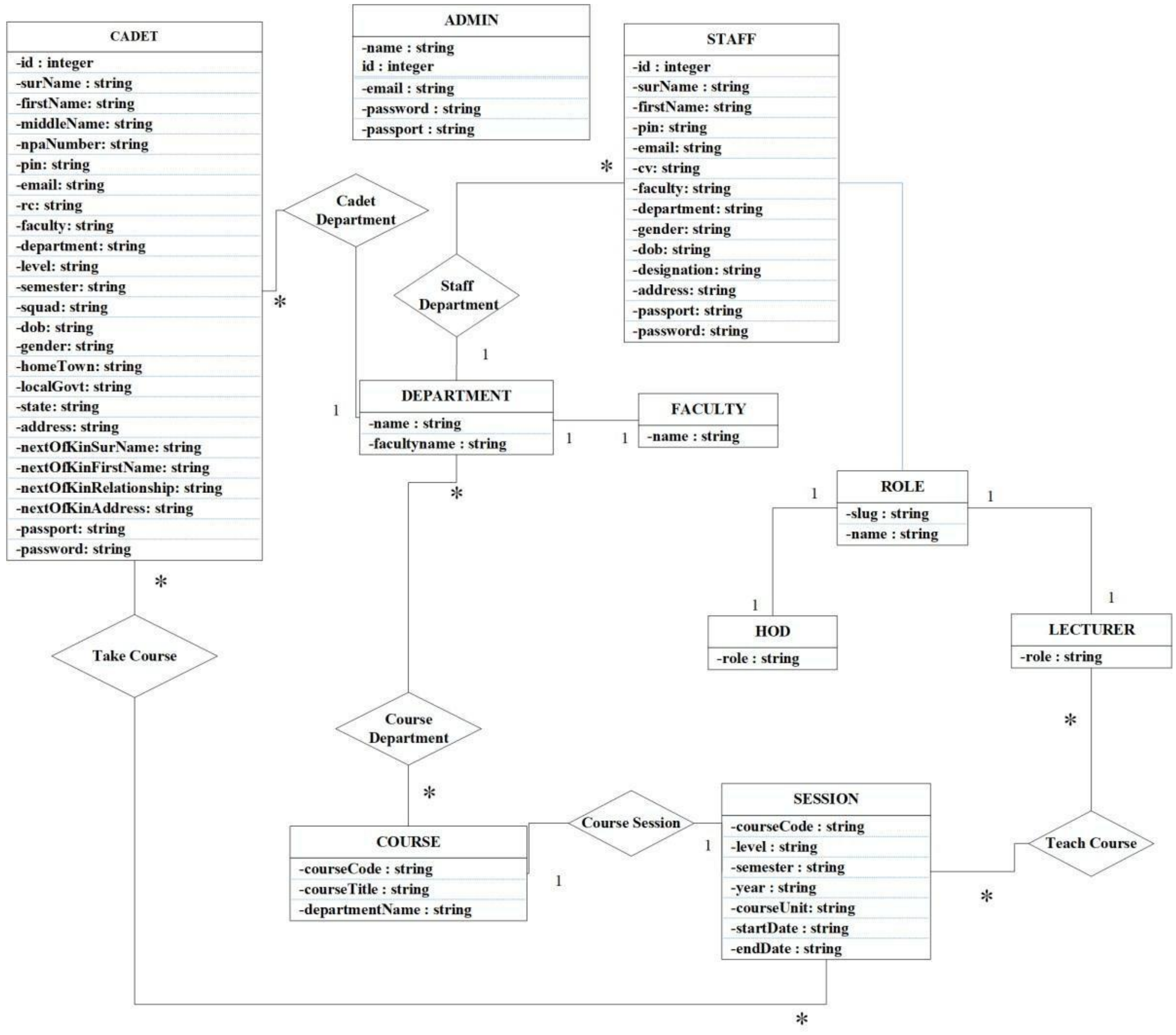

Figure 3. 9: E-R Diagram for Secure Student Information Management System

### 3.5.3 User Interface

This involves the design of all the interface the user need to interact with the system.

#### 3.5.3.1 Input Design Form

The Figure below shows the interface of the login page, where student can sign in to access the student portal.

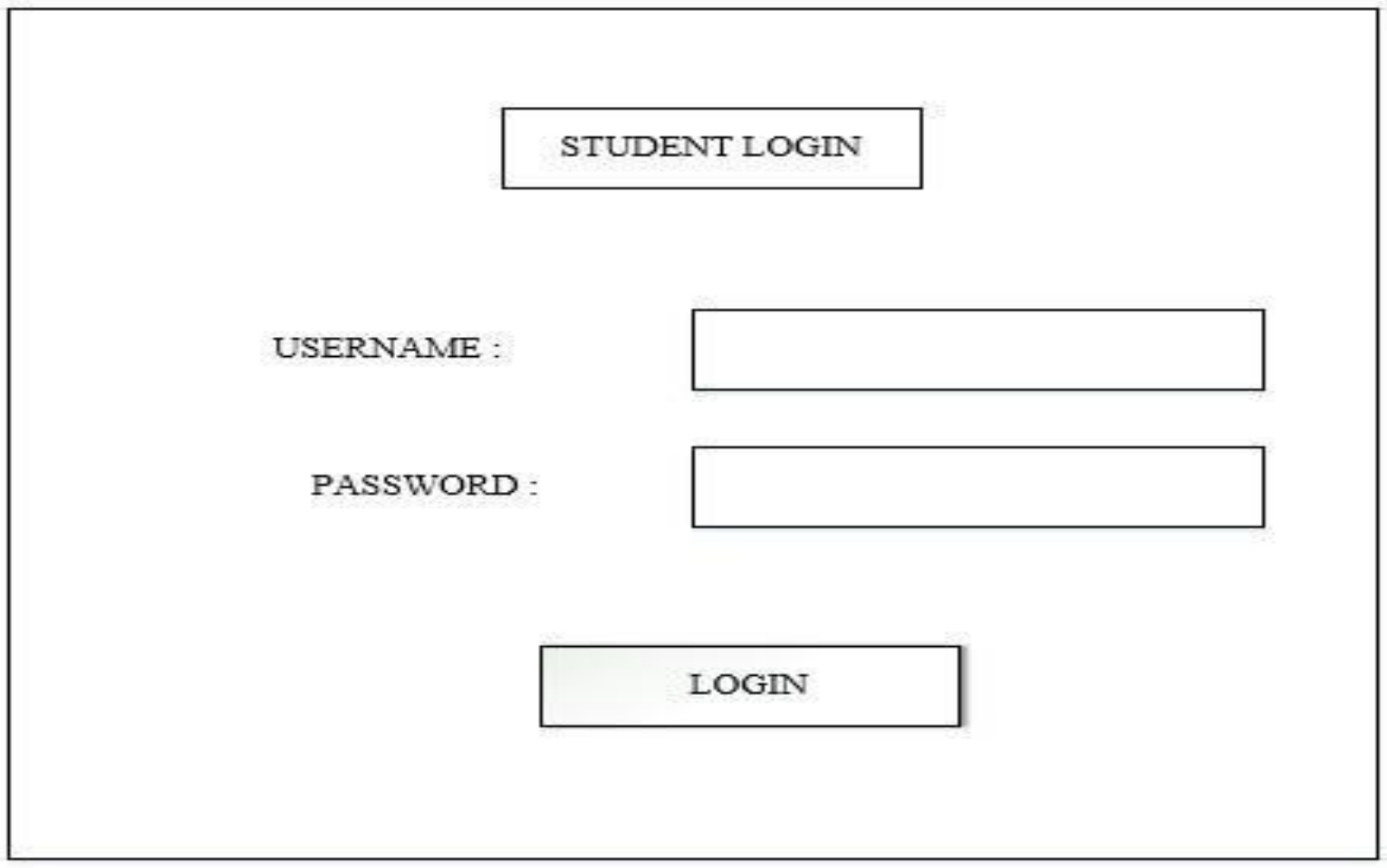

Figure 3. 10: Input Form for Student Login

The Figure below shows the interface for student registration.

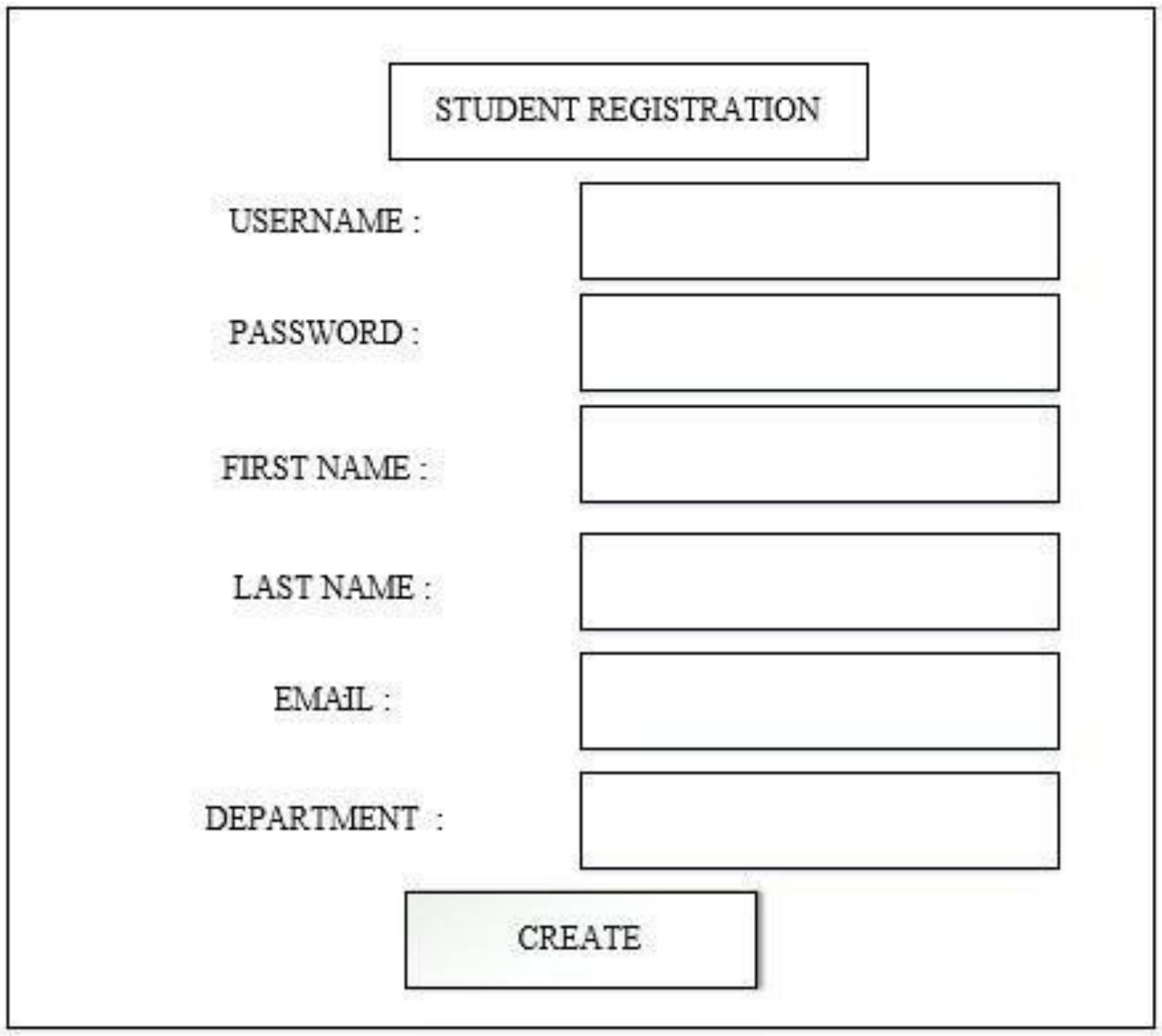

Figure 3. 11: Input Form for Student Registration

# CHAPTER FOUR

# SYSTEM IMPLEMENTATION AND TESTING

## 4.1 Introduction

This chapter comprises of the detail implementation of the system as concern with the preparation of resources i.e. hardware and software that are required for effective functionality of a newly designed system, testing these resources to ensure that they meet the designed objective and eventually change over to the new system.

## 4.2 Choice and Justification of Programming Language

The Choice of the programming technologies languages used in the implementation of this project was determine through the design phase. The result of the design phase makes it possible to represent the system as a set of interacting object which allow the server side of the system to be implemented using PHP (Hypertext Pre-processor).

PHP is a server-side scripting language designed for web development. As of December 2017, PHP makes up over 83% of server side languages used on the internet and was originally created by Rasmus Lerdorf in 1995, the reference implementation of PHP is now produced by The PHP Group. While PHP originally stood for Personal Home Page, it now stands for PHP: Hypertext Pre-processor, a recursive back acronym. The current version of PHP is PHP 7.2

PHP, known as the most popular server-side scripting language in the world, has evolved a lot since the first inline code snippets appeared in static HTML files. A lot of Web application framework exist for PHP used in the development of web application such as Laravel, Symfony, CodeIgniter, Yii2, Phalcon, CakePHP, Zend, Slim, FuelPHP, PHPixie e.t.c. Web application framework facilitate and streamline the process of backend web development. The PHP Framework used in this project was Laravel 5.5 to ensure the security of the project.

Laravel PHP Framework is very popular for custom software development. It is the Most Starred PHP Framework on Github with more than 35,000 developers from all over the world (mostly from the USA). Laravel is an outstanding member of a new generation of web frameworks. Laravel was created by Taylor Otwell in 2011.

Laravel provide some features which make it suitable for the implementation of the system server side such as

i. Frontend: Laravel use blade template engine which allow developers to perform template inheritance, sections and secure access to data.

ii. Security: Laravel consists of the various security features which includes the following

   a. Authentication: Laravel makes implementing authentication very simple, which are made up of guards and providers to define how users are authenticated for each request.

   b. Encryption: Laravel's encrypter uses OpenSSL to provide AES-256 and AES-128 encryption algorithm. All Laravel encrypted values are signed using a message authentication code (MAC) so that their underlying value cannot be modified once encrypted. i.e Credit Card details e.t.c.

   c. Hashing: Laravel Hash facade provide secure Bcrypt and Argon2 hashing for storing user passwords.

d. Authorization: Laravel provides a simple way to authorize user actions against a given resource.

e. API Authentication: Laravel makes API authentication a breeze using Laravel Passport, which provide a full OAuth2 server implementation for Laravel application in a matter of minute.

f. Password Reset.

Laravel also helps to secure the web application by protecting it against the most serious security risks: SQL injection, cross-site request forgery and cross-site scripting.

i. Error and Exception Handling.

ii. Routing.

iii. MVC pattern. iv. Caching.

v. Simplified Database interaction: Laravel makes interacting with databases extremely simple across a variety of database backends using raw SQL, fluent query builder, and the Eloquent ORM. (Ali, 2018)

For a robust development several languages were used for the implementation of the client side which includes:

i. Laravel Blade Template Engine.

ii. HTML 5.

iii. Bootstrap Framework v4.0.0-alpha.6 built on Cascading Style Sheet (CSS).

iv. Vue Js Framework built on JavaScript and npm package manager for installation of other Front-end libraries.

## 4.3 Deployment Platform

The client code is deployed on windows 10 operating system and the web service code is deployed on Laragon a localhost webserver that consist of Apache server. “Laragon is the best and fastest local server by far” SniffleValve. Laragon v.3.1.6 localhost webserver which consist of Apache/2.4.27 server (Win64), OpenSSL/1.0.2l , PHP version: 7.1.7 and phpMyAdmin built on MariaDB was used for the MySQL database management Interface. (Valve, 2018).

## 4.4 Software Testing

The unit testing approach was adopted in testing the codes written. The procedure adopted for the unit test is;

i. The module interface is tested to ensure that information properly flows into and out of the program unit under test.

ii. The local data structure is examined to ensure that data stored temporarily maintained its integrity during all steps in an algorithm’s execution.

iii. All the statements are executed at least once and error handling paths are tested.

## 4.5 Documentation

Documentation is very important in the development of any software application. This is because documentation makes the software application easier to all users, and if an application is not well documented it becomes difficult to use.

### 4.5.1 Hardware Requirement

The software application designed needed the following hardware for effective operation of the newly designed system:

i. Pentium IV PC / Laptop.

ii. The Random Access Memory (RAM) should be at least 2 GB.

iii. Enhanced Keyboard iv. At least 100 GB hard Disk.

v. E.G.A/V.G.A, a coloured Monitor.

vi. An uninterruptible Power Supply (UPS) Unit vii. Voltage Stabilizer: This facilitates the regulation of voltage needed by the computer system in order to avoid electrical damage of the system.

### 4.5.2 Software Requirement

The software requirements for this system include:

i. Windows 7/8/8.1/10 or MacOSx Operating system.

ii. Laragon v 3.

iii. Visual Code Studio.

## 4.6 System Testing

This is the process of testing the web application, this testing was done using Google Chrome web browser. Figure 4.1 – 4.31 show the testing of this project.

i. Admin login

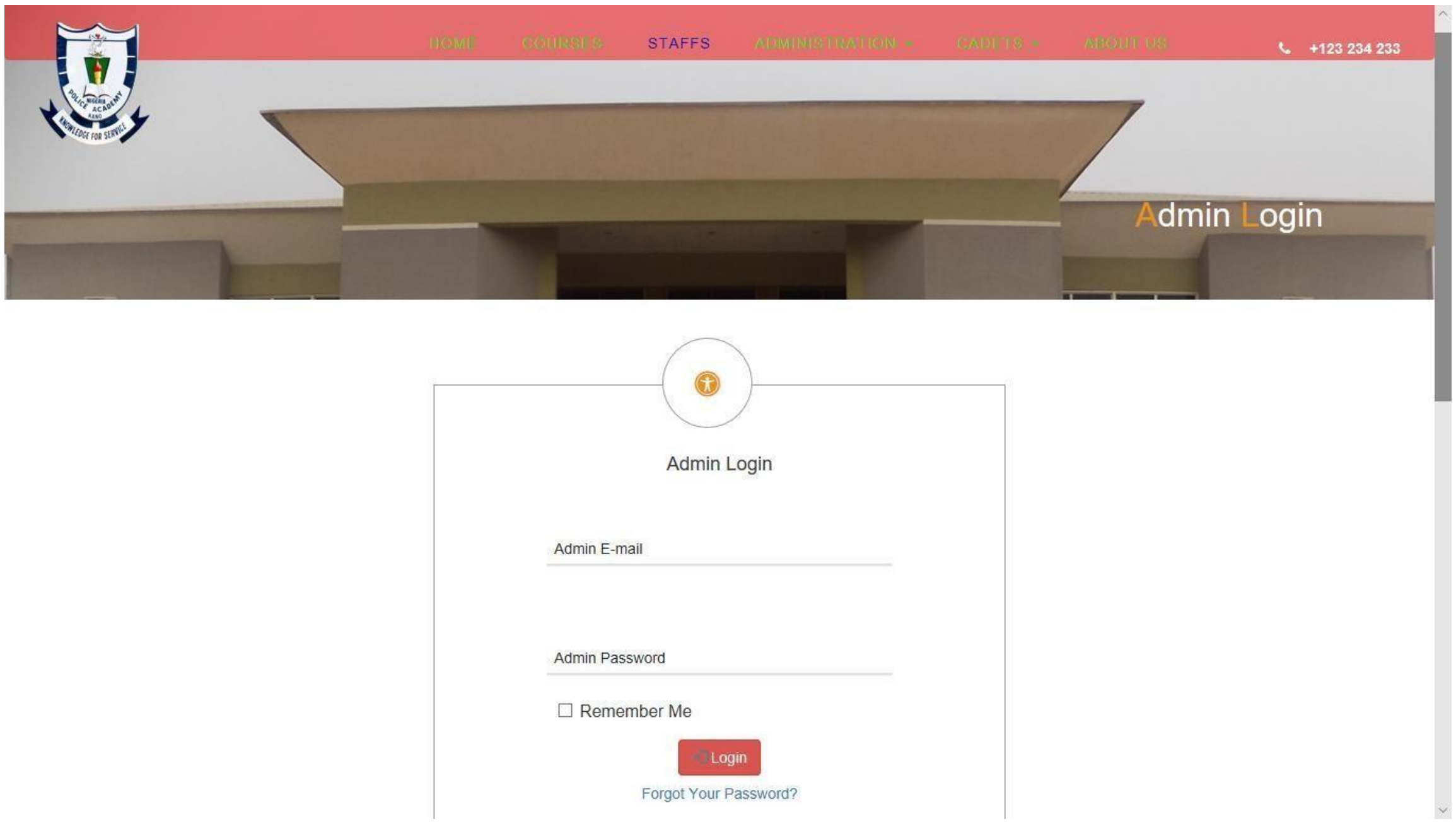

Figure 4. 1 Admin Login Page.

ii. Admin dashboard

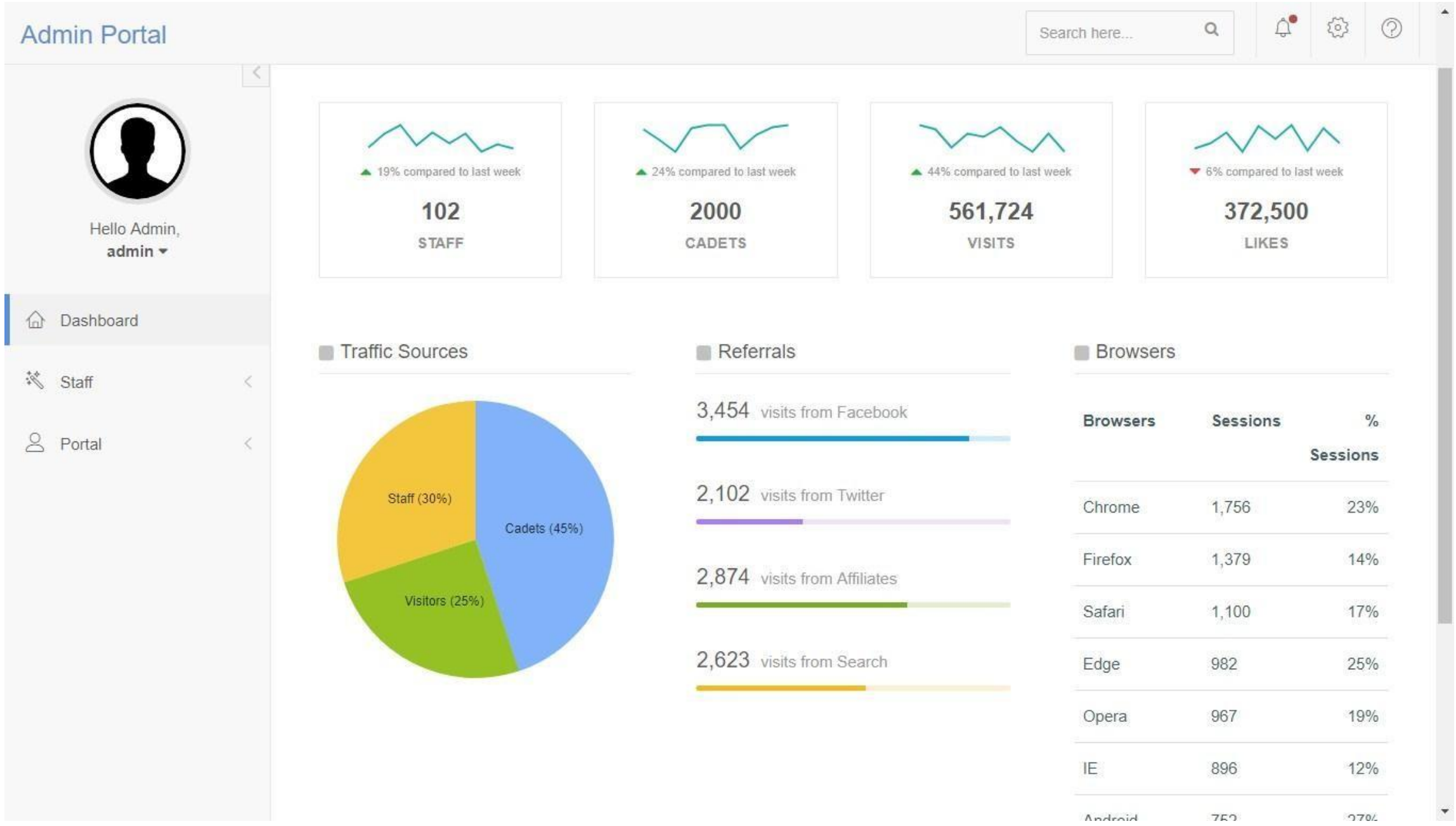

Figure 4. 2 Admin Dashboard

iii. Admin view staff

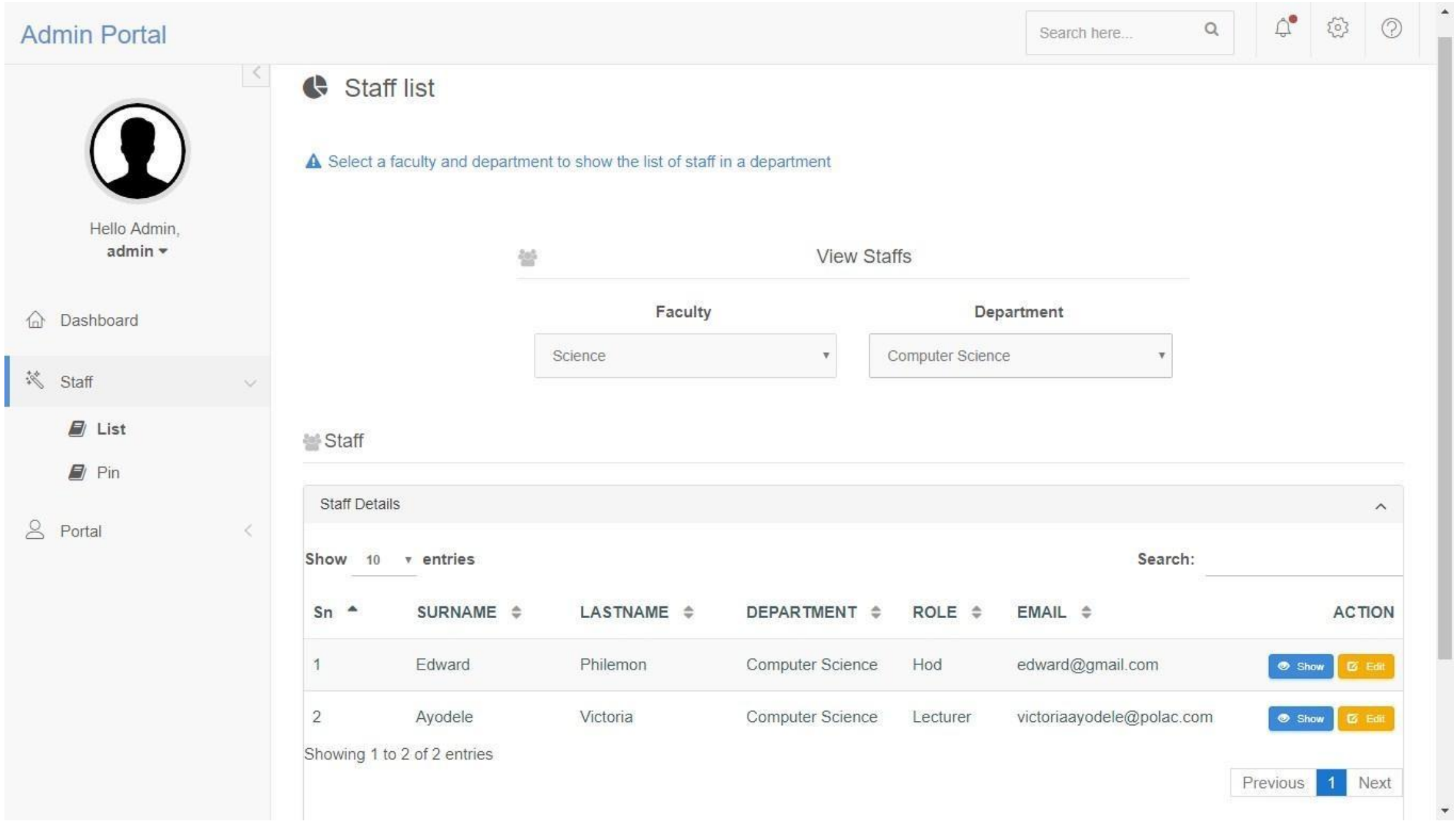

Figure 4. 3 Admin view staff.

iv. Admin generate staff registration pin

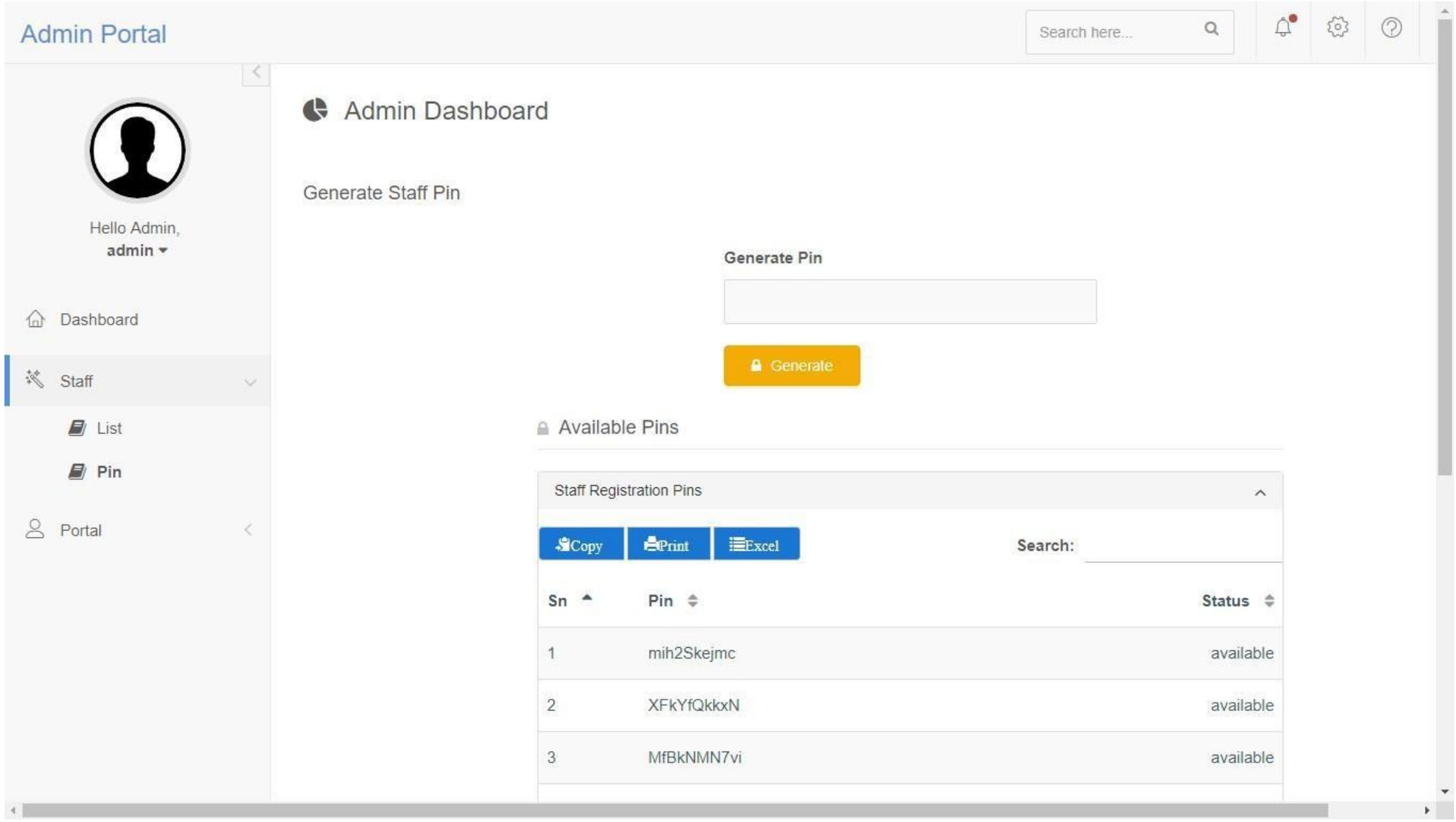

Figure 4. 4 Admin create staff registration pin.

v. Admin create event

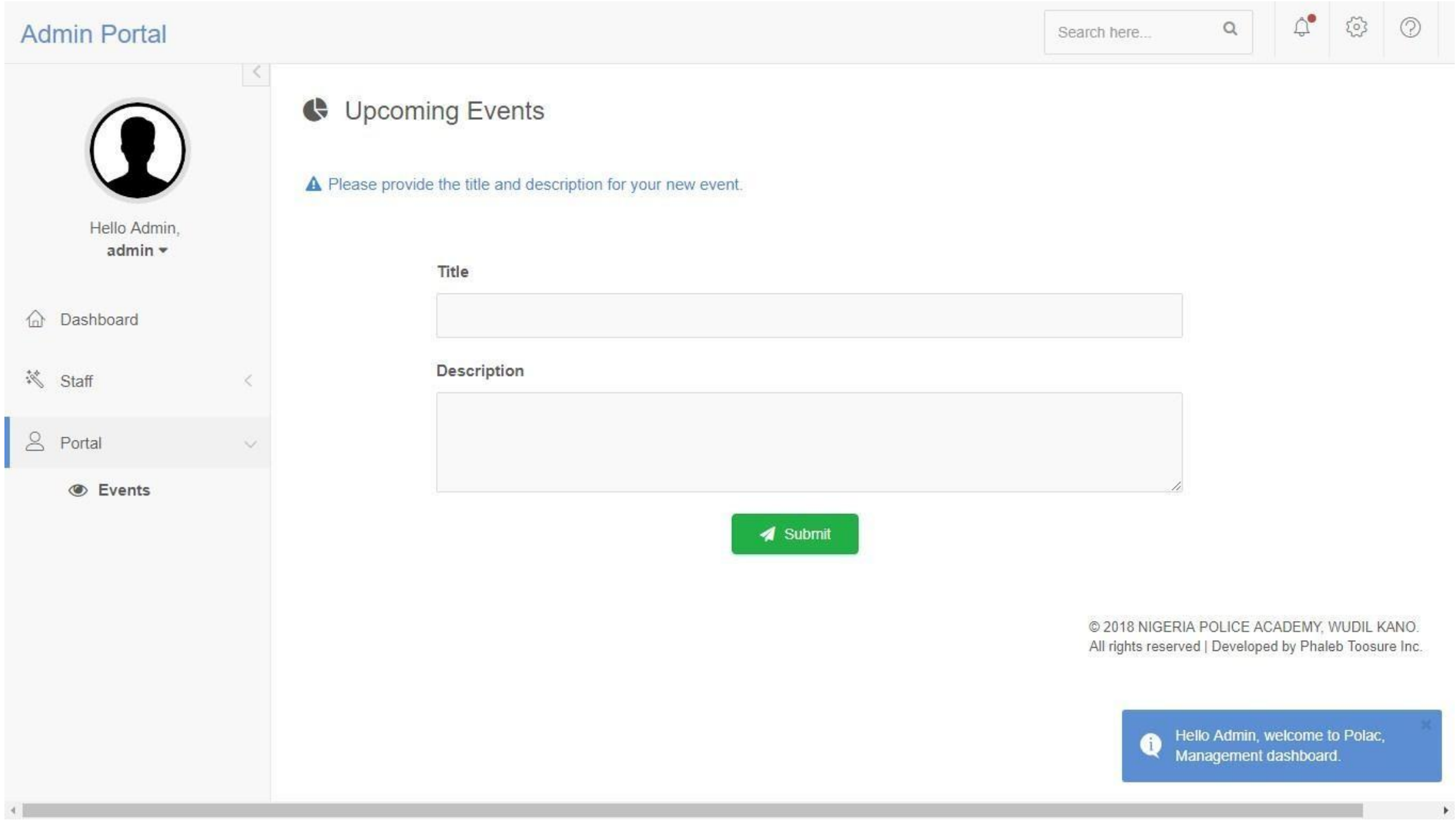

Figure 4. 5 Admin create event

vi. Admin edit staff data

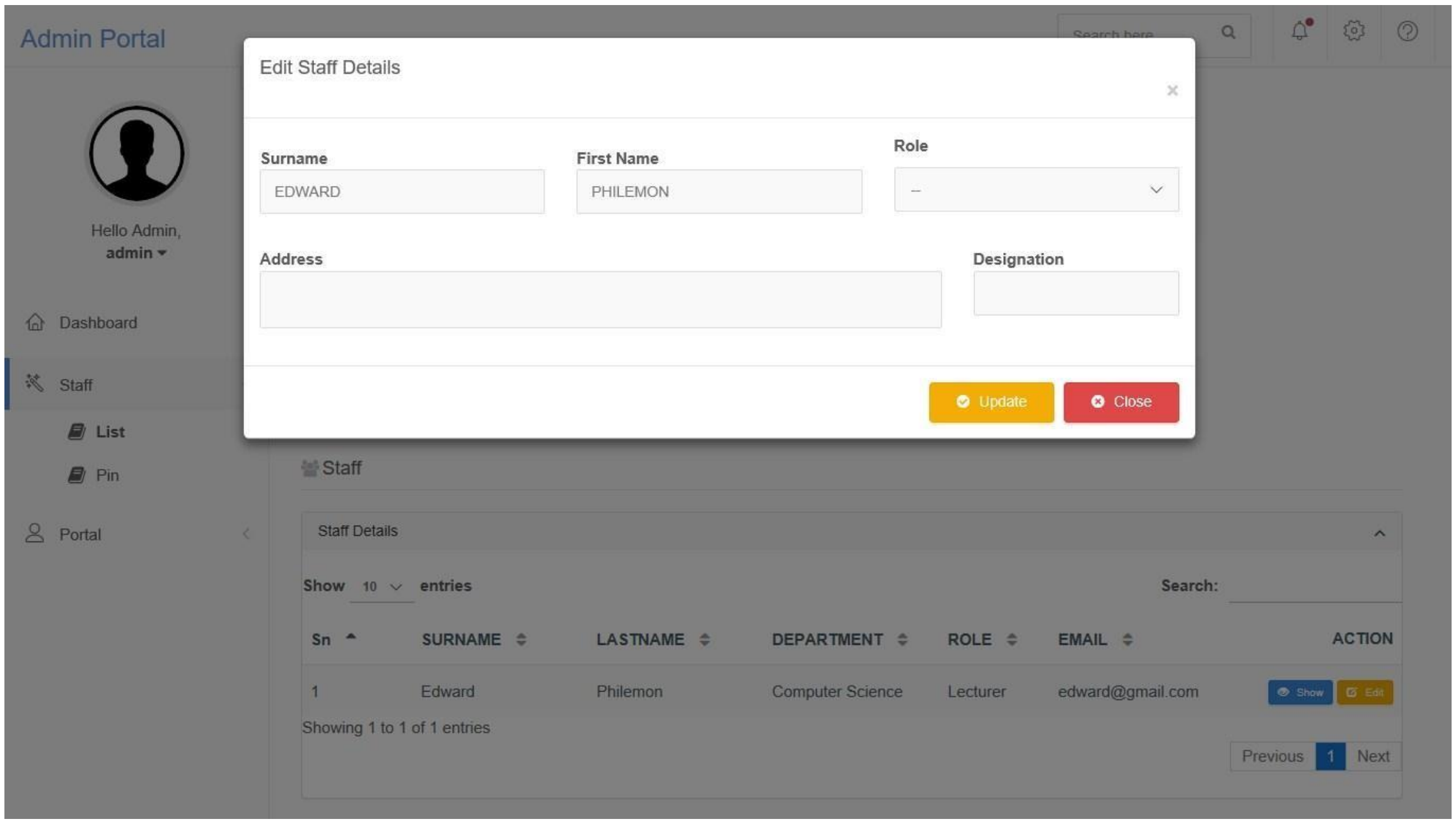

Figure 4. 6 Admin edit staff data.

vii. Staff registration

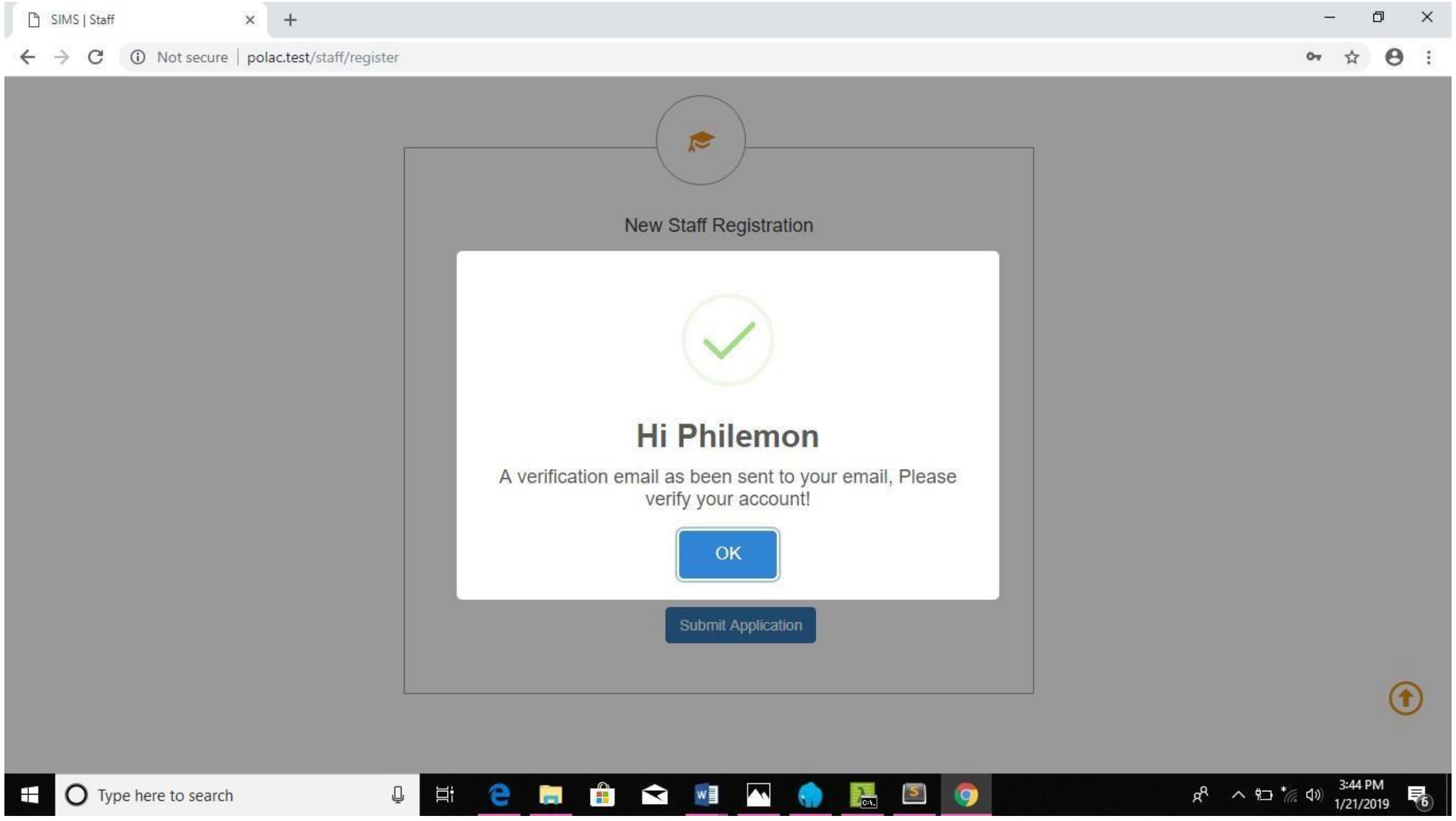

Figure 4. 7 Staff registration

viii. Staff login

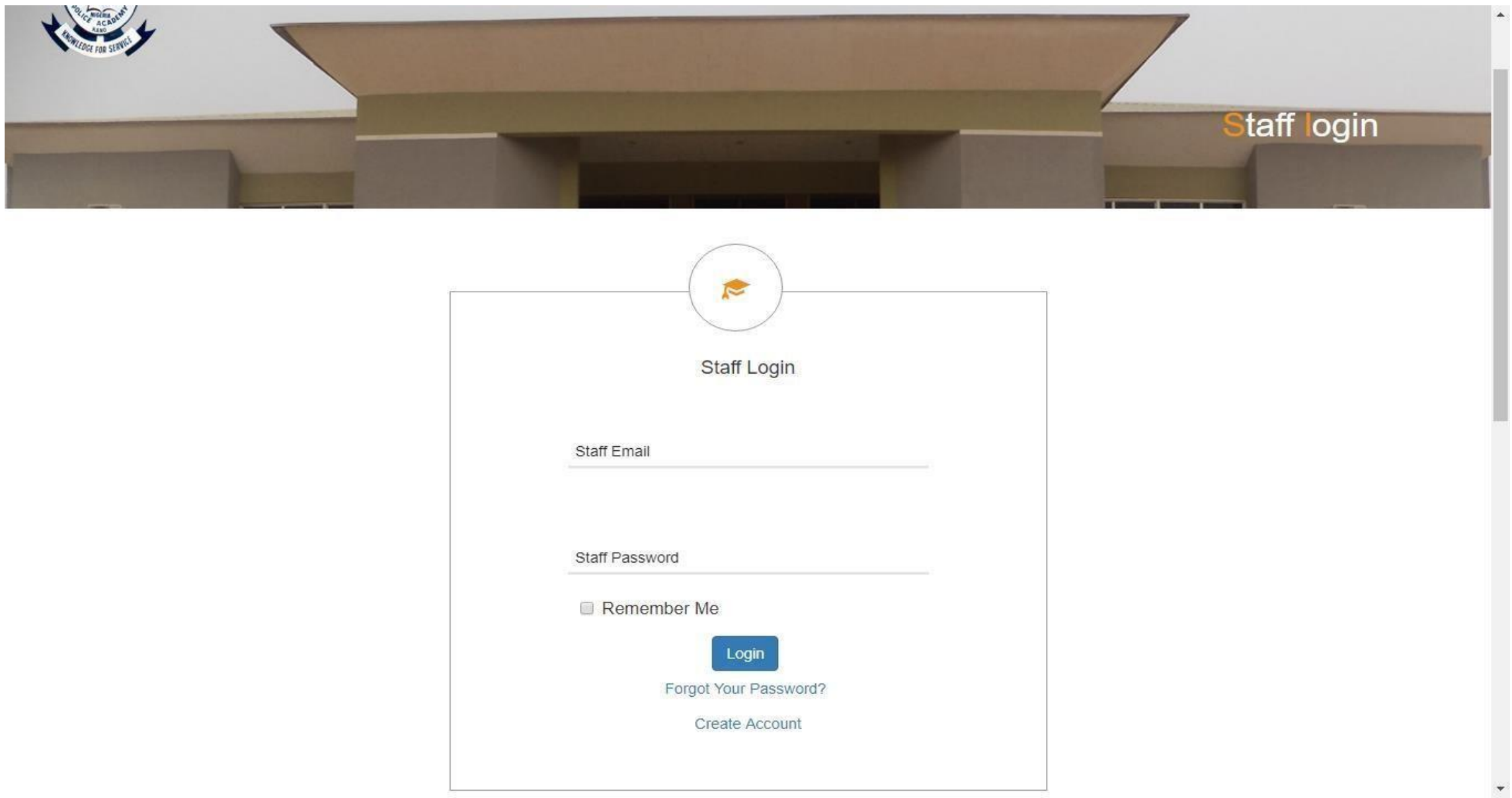

Figure 4. 8 Staff login.

ix. Hod dashboard

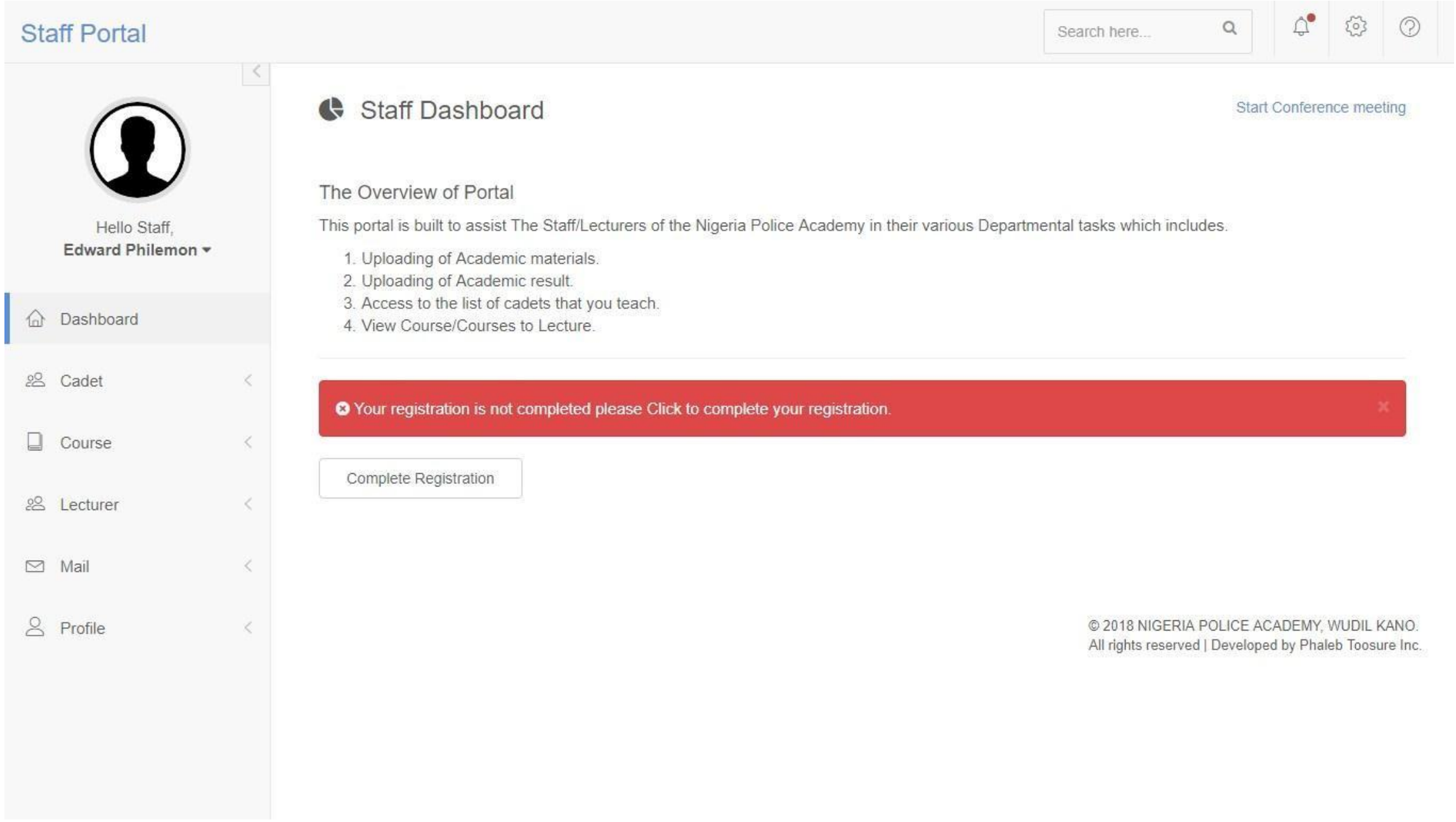

Figure 4. 9 Hod Dashboard

x. Hod roles

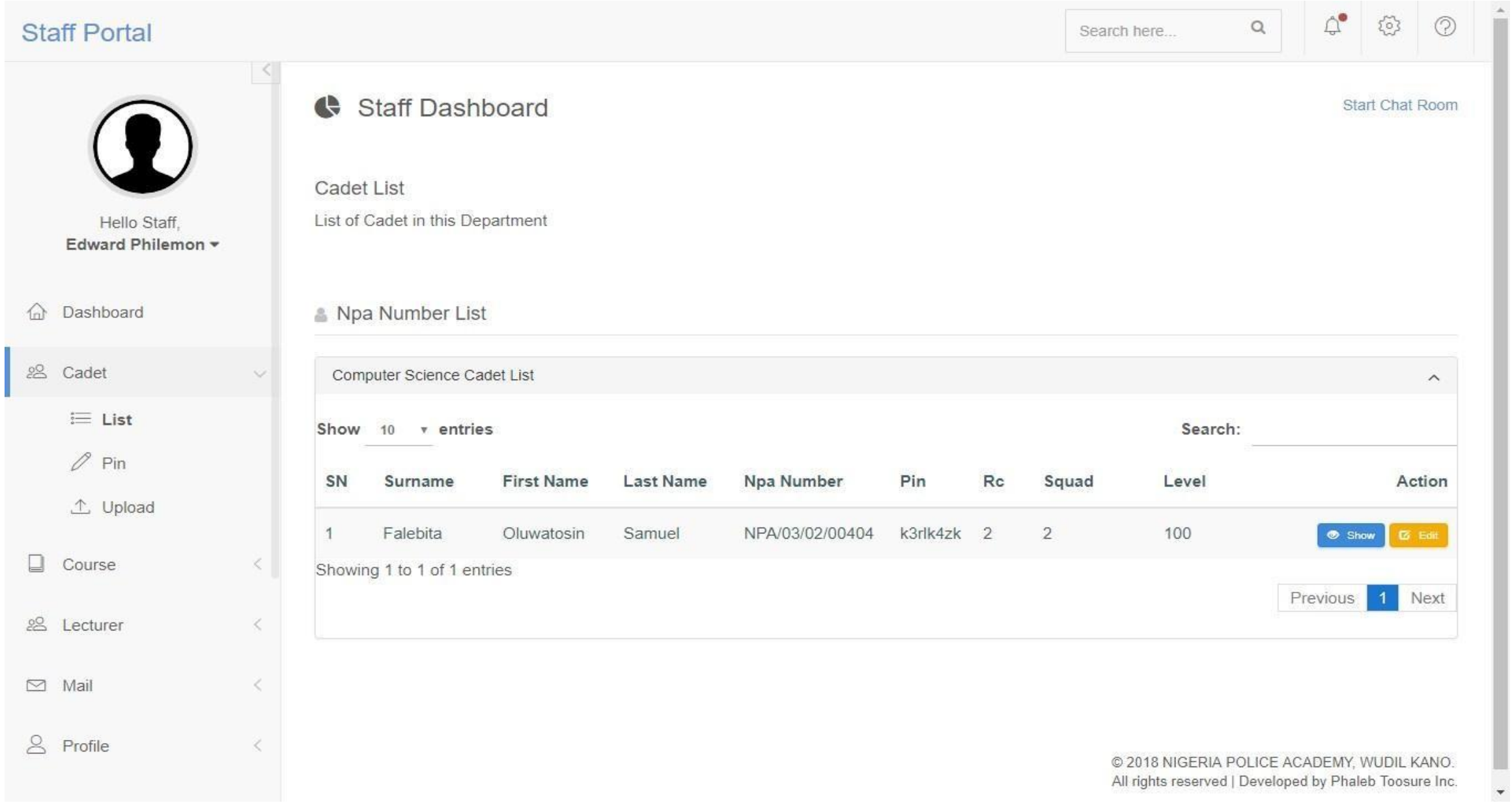

Figure 4. 10 Cadet list, edit

xi. Generate Cadet Pin

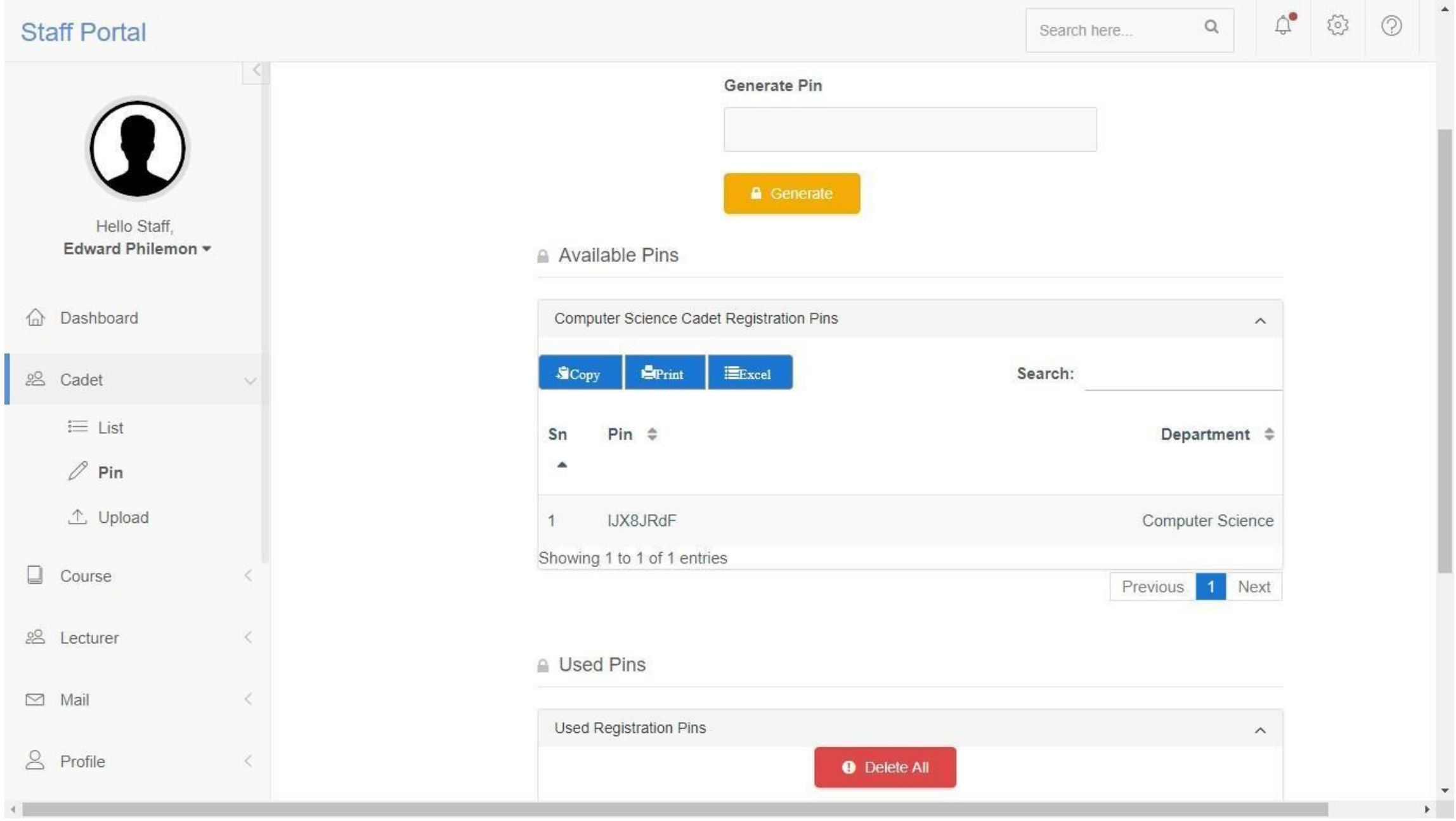

Figure 4. 11 Create cadet registration pin

xii. Hod Upload Npa Number.

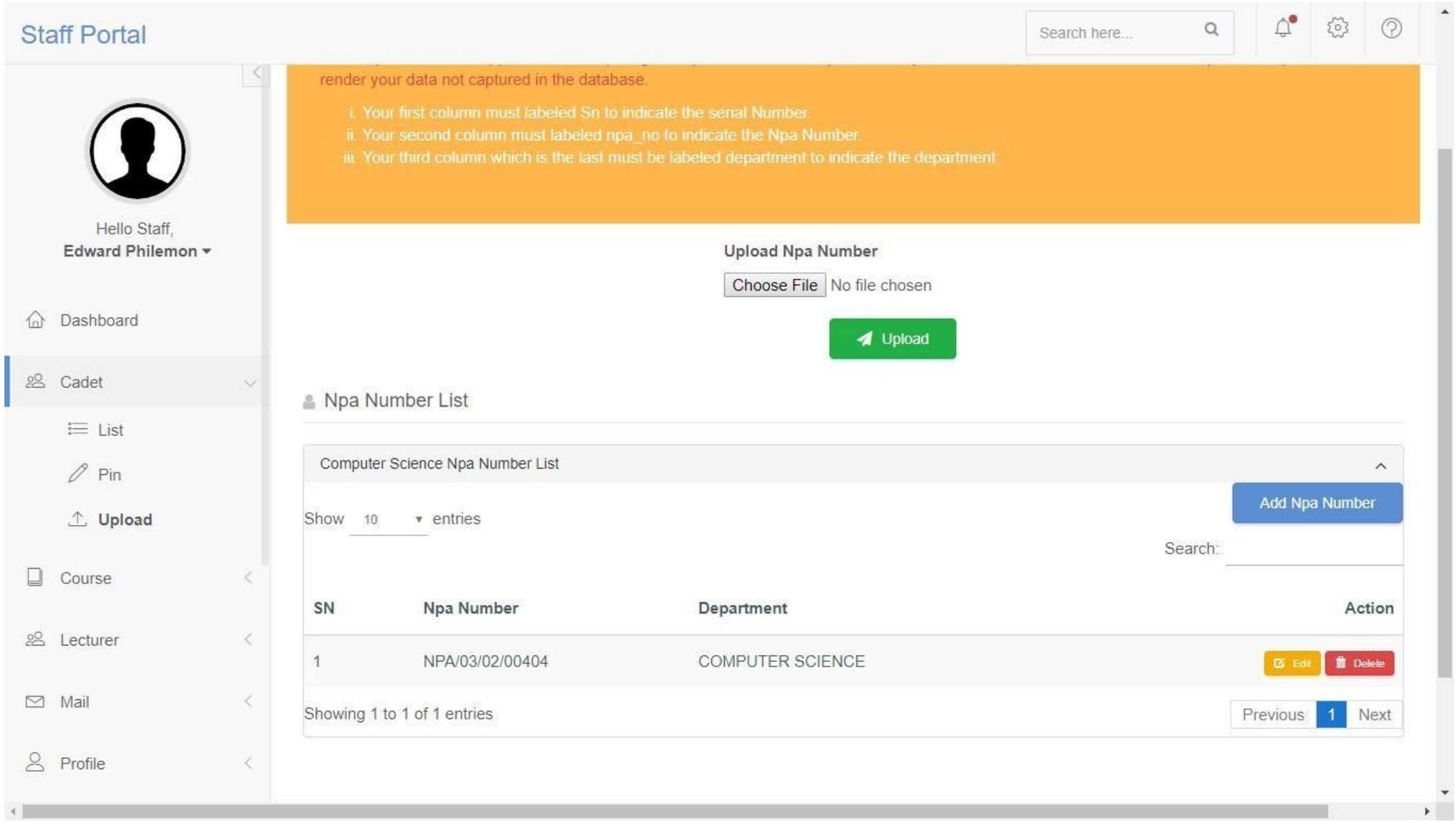

Figure 4. 12 Upload cadet npa number.

xiii. Hod create Departmental Courses.

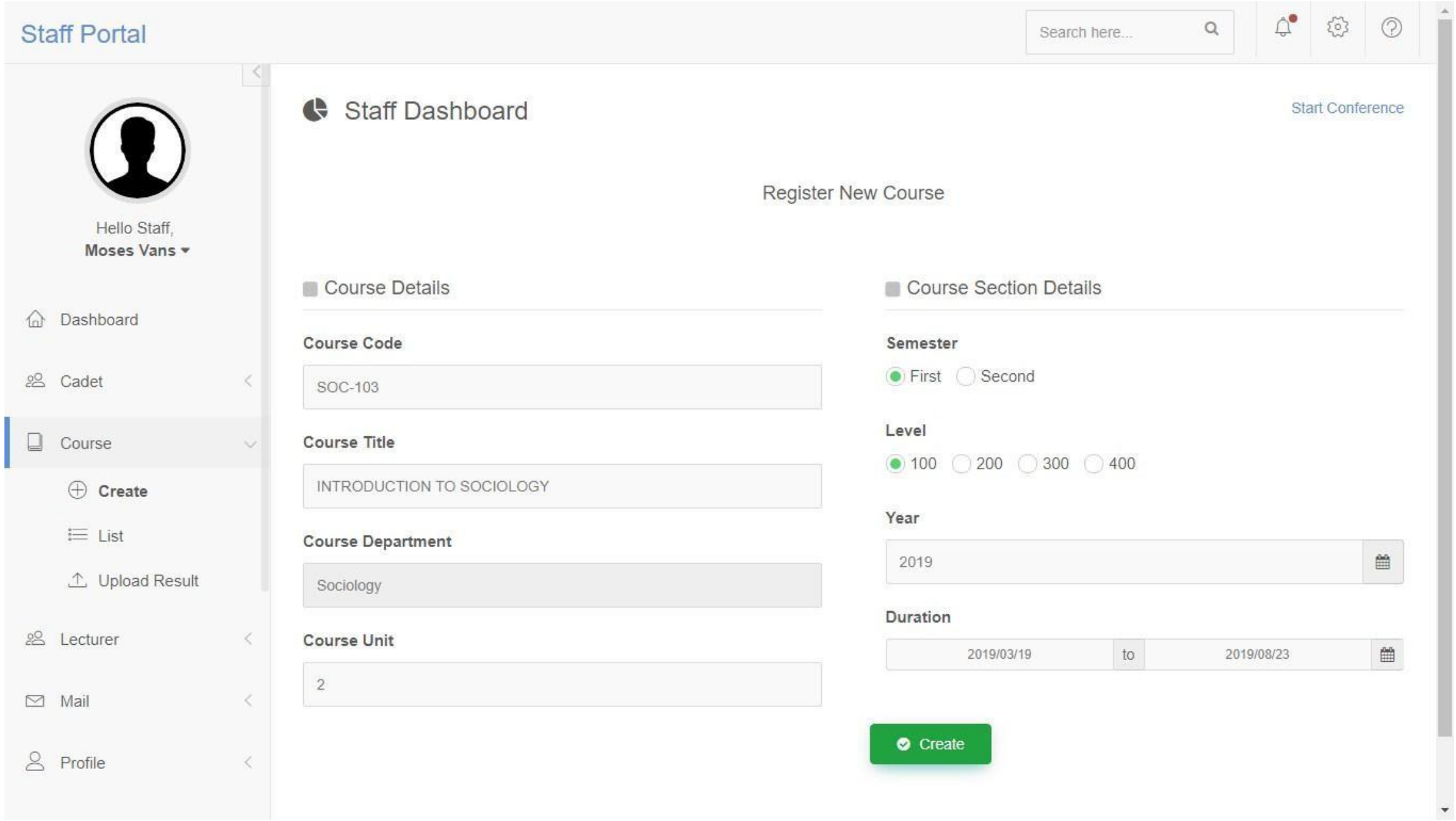

Figure 4. 13 Create course.

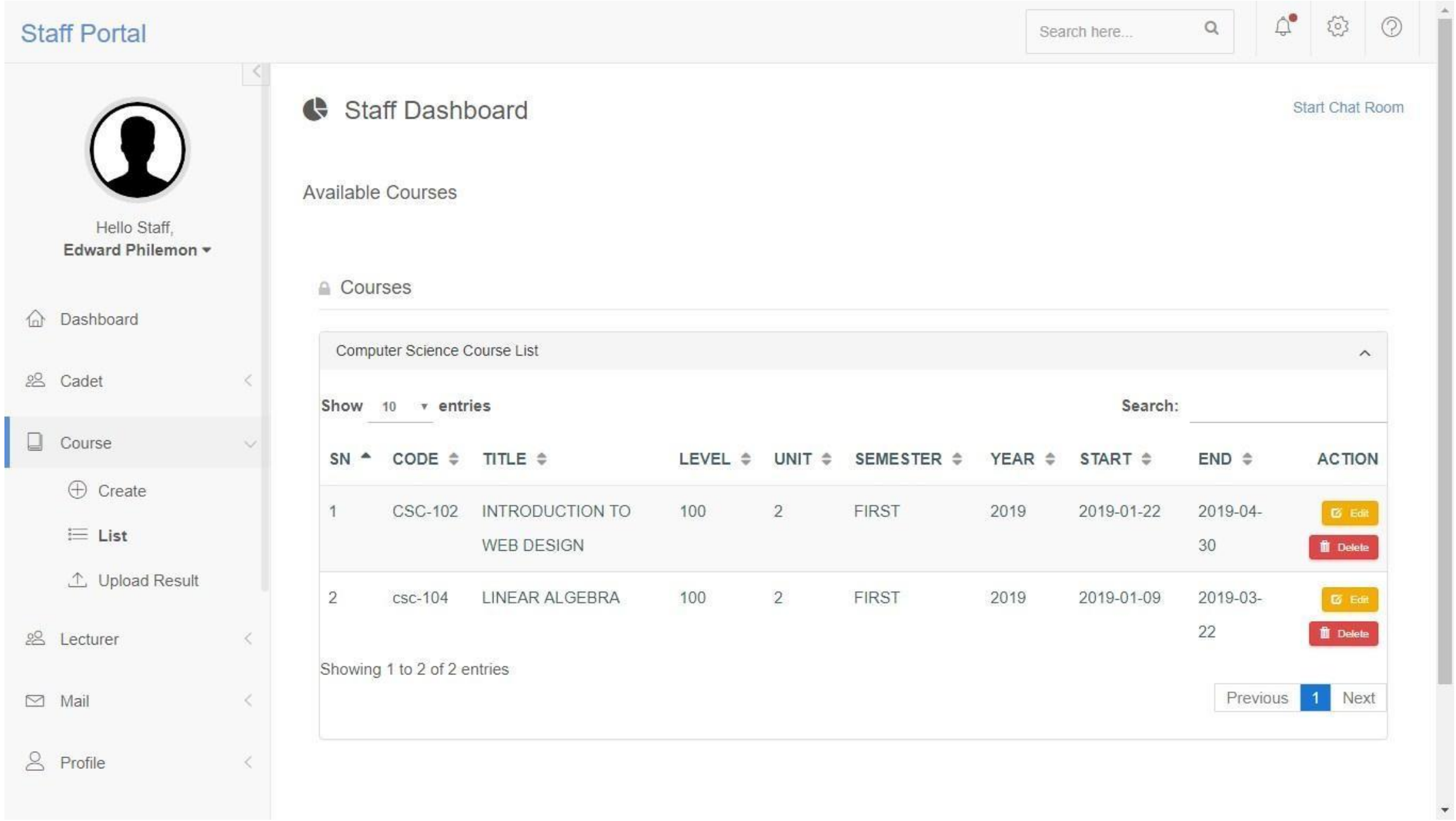

Figure 4. 14 Course list, edit and delete.

xiv. Hod assign Courses to Lecturer

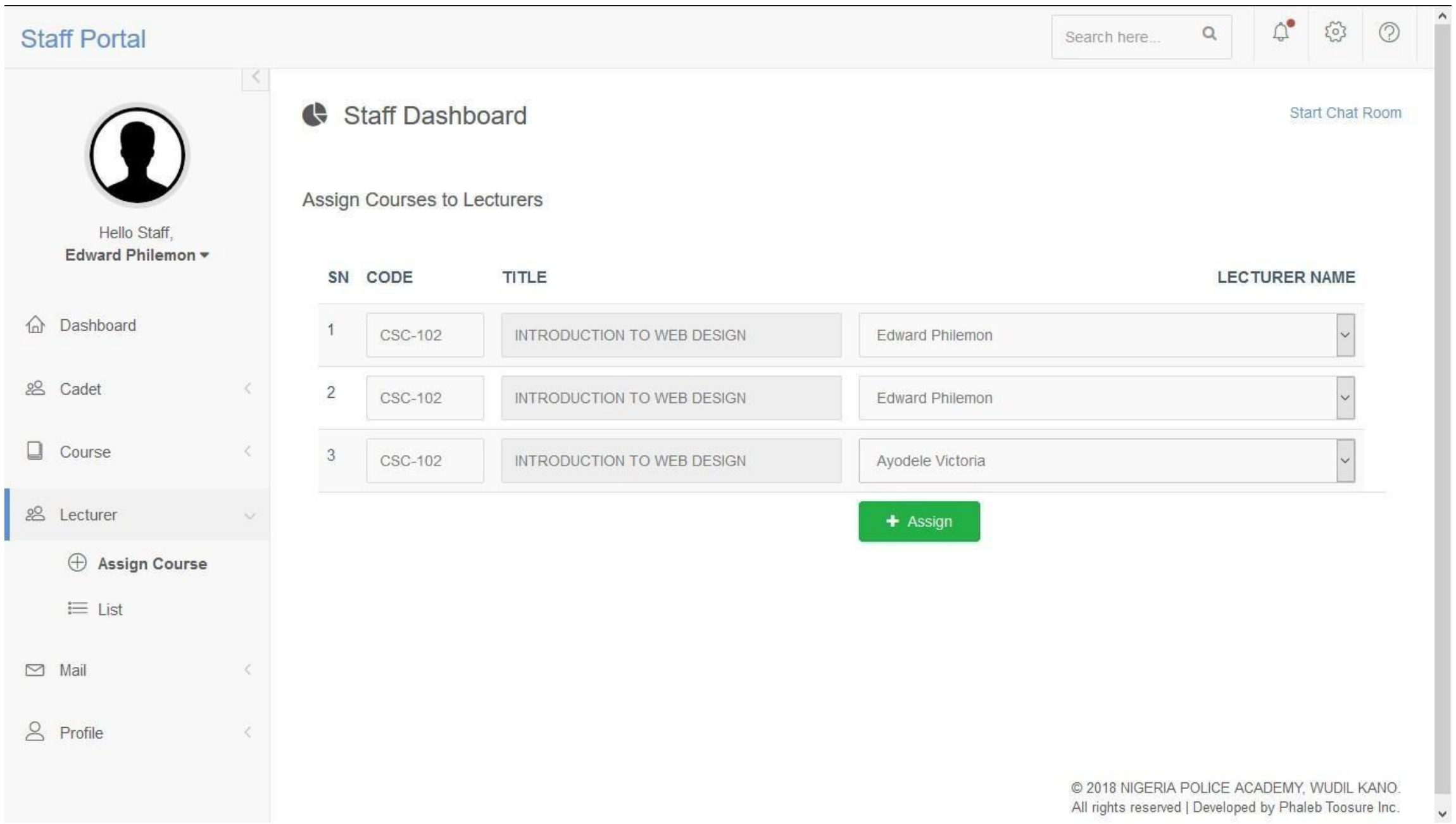

Figure 4. 15 Assign course to lecturers.

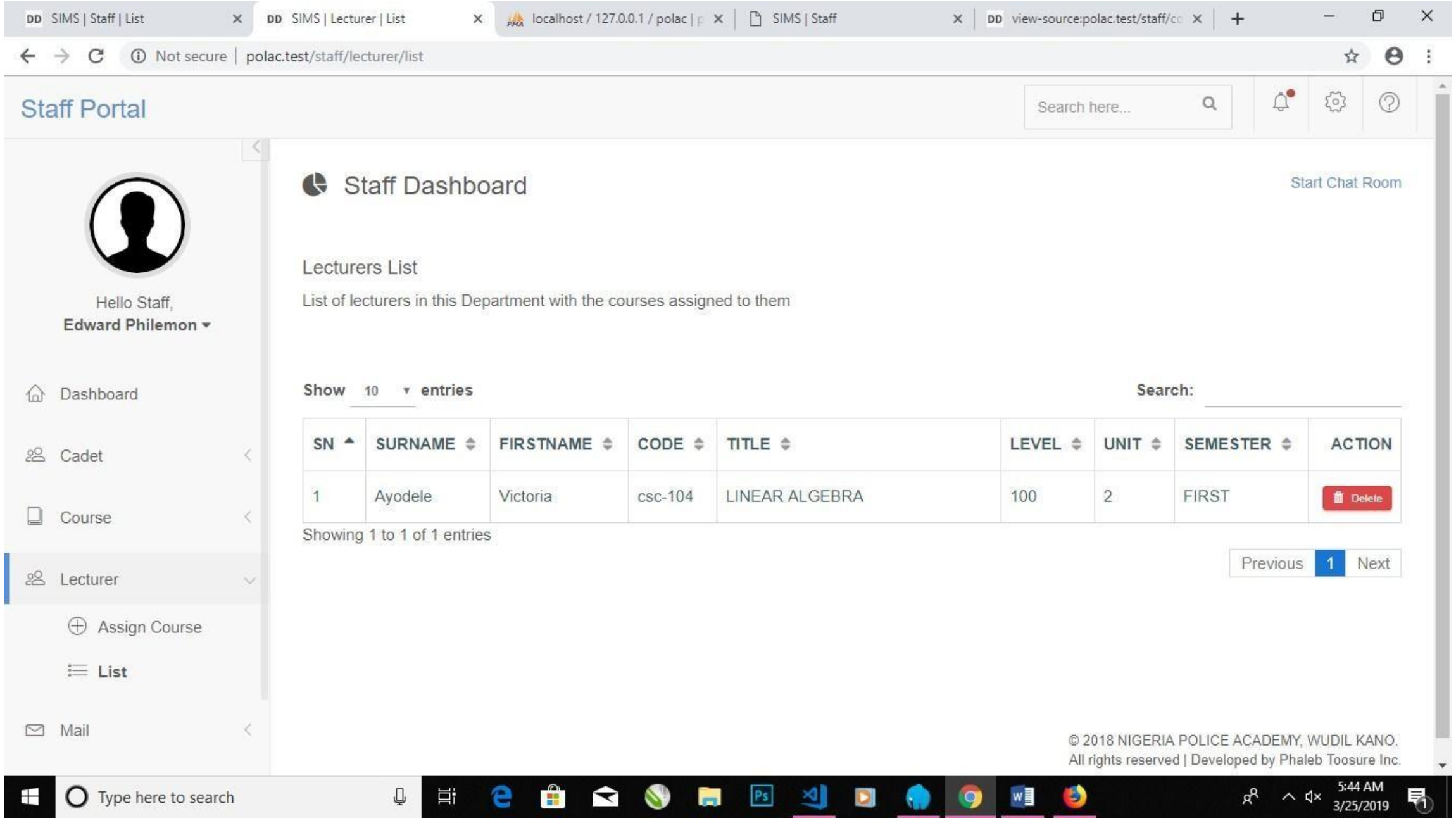

Figure 4. 16 List of lecturers.

xv. Hod Upload Cadet Results

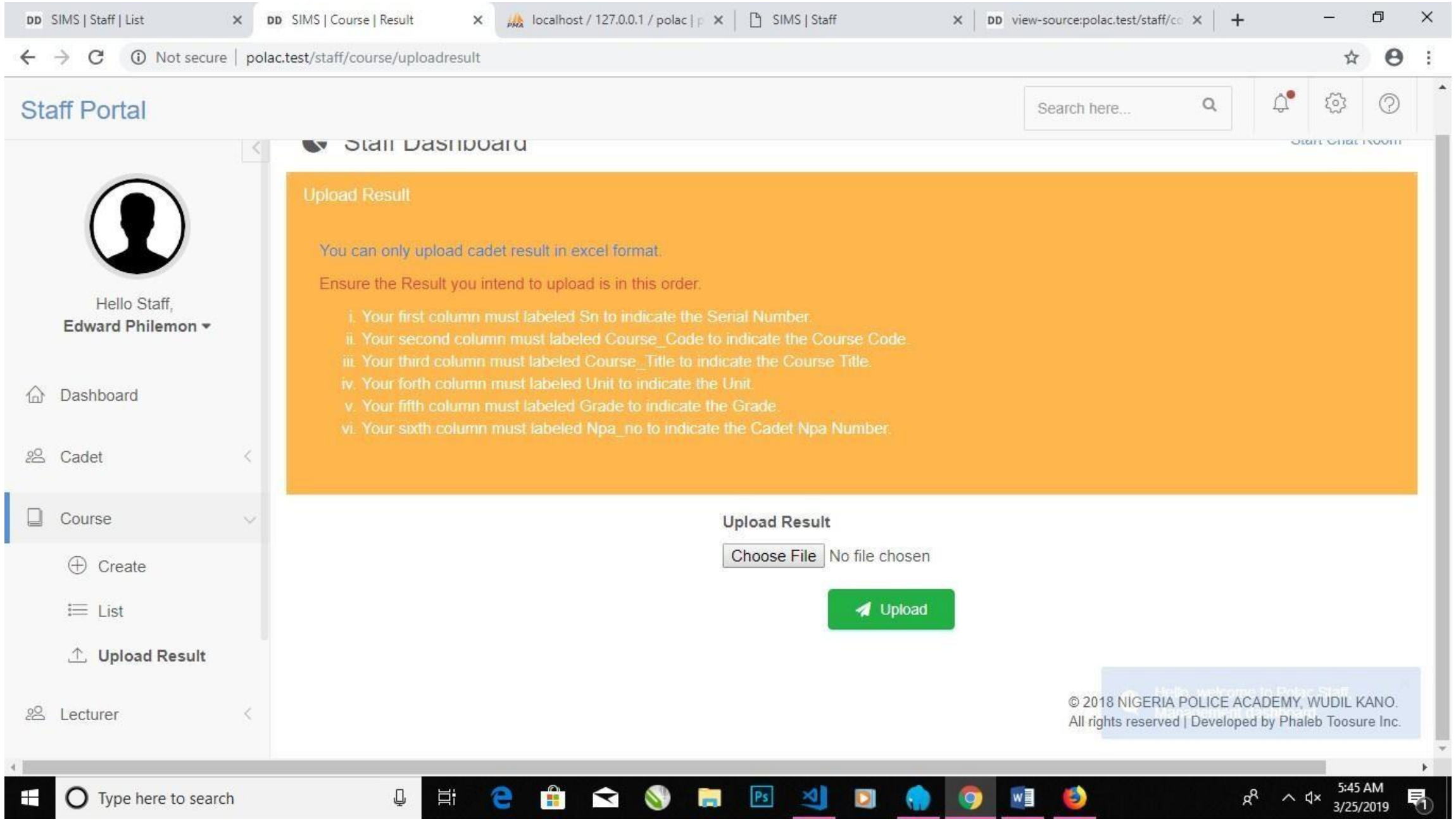

Figure 4. 17 Upload results.

xvi. Hod Profile

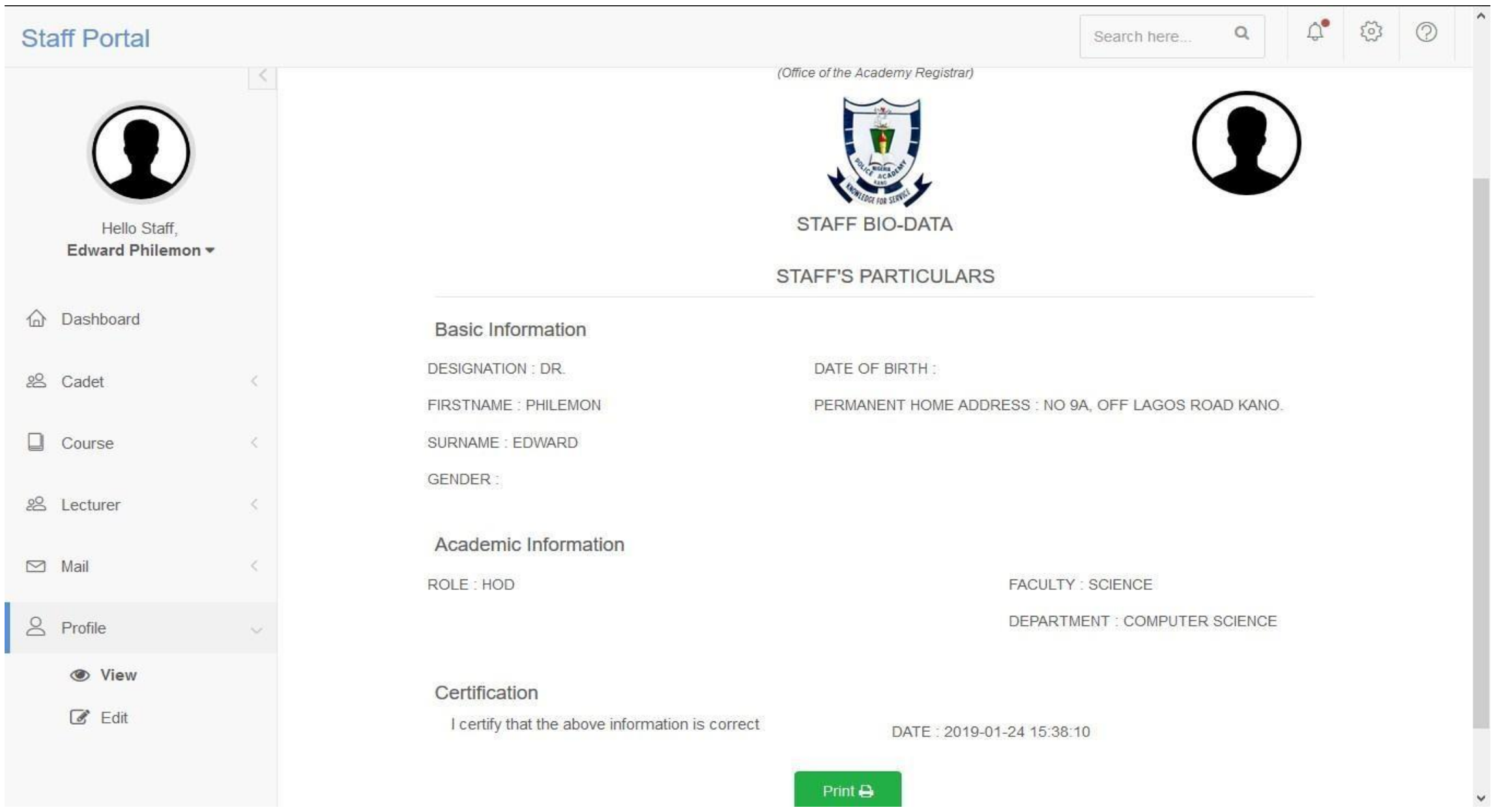

Figure 4. 18 Hod Profile.

xvii. Hod Update Profile.

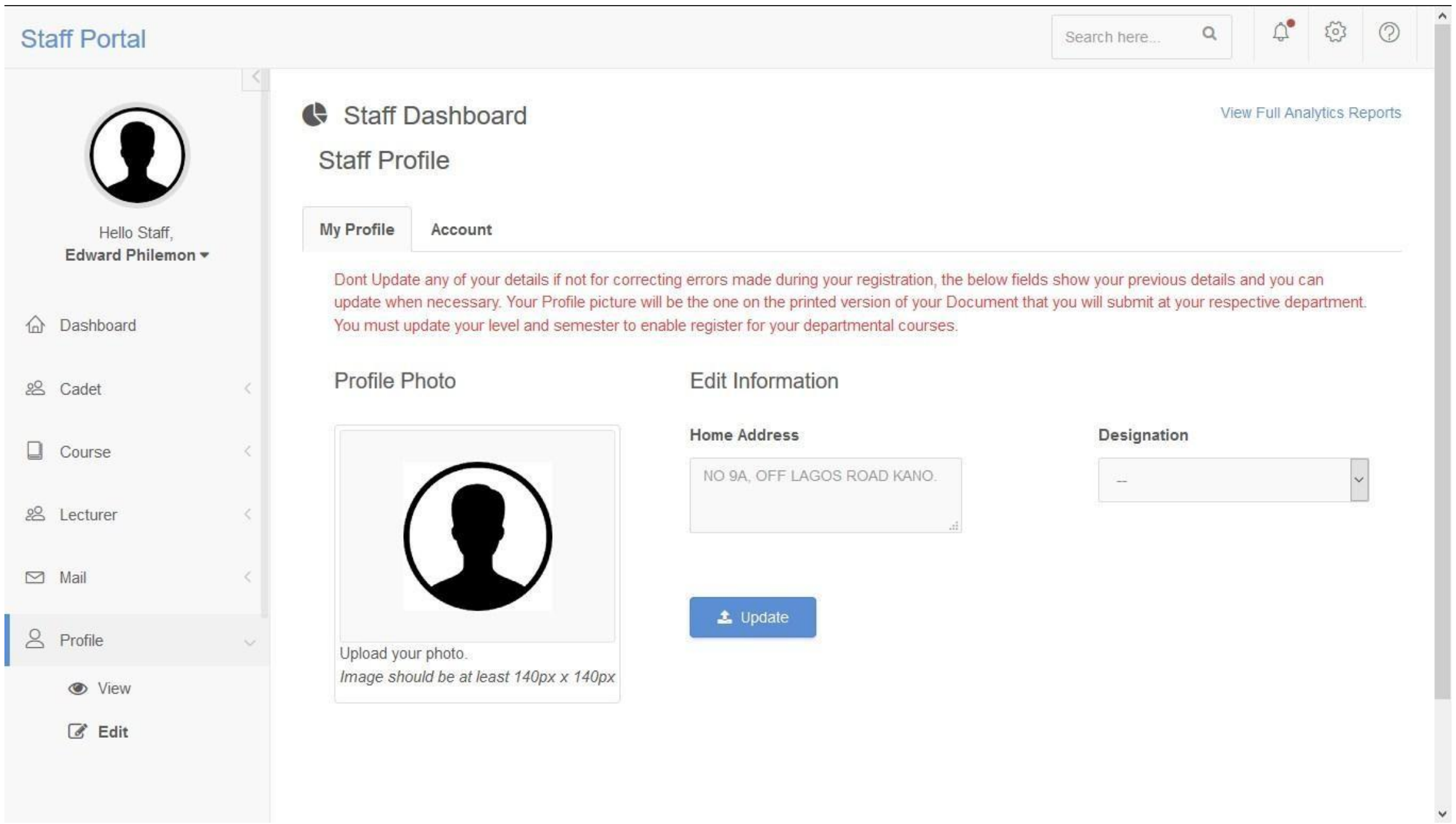

Figure 4. 19 Update profile.

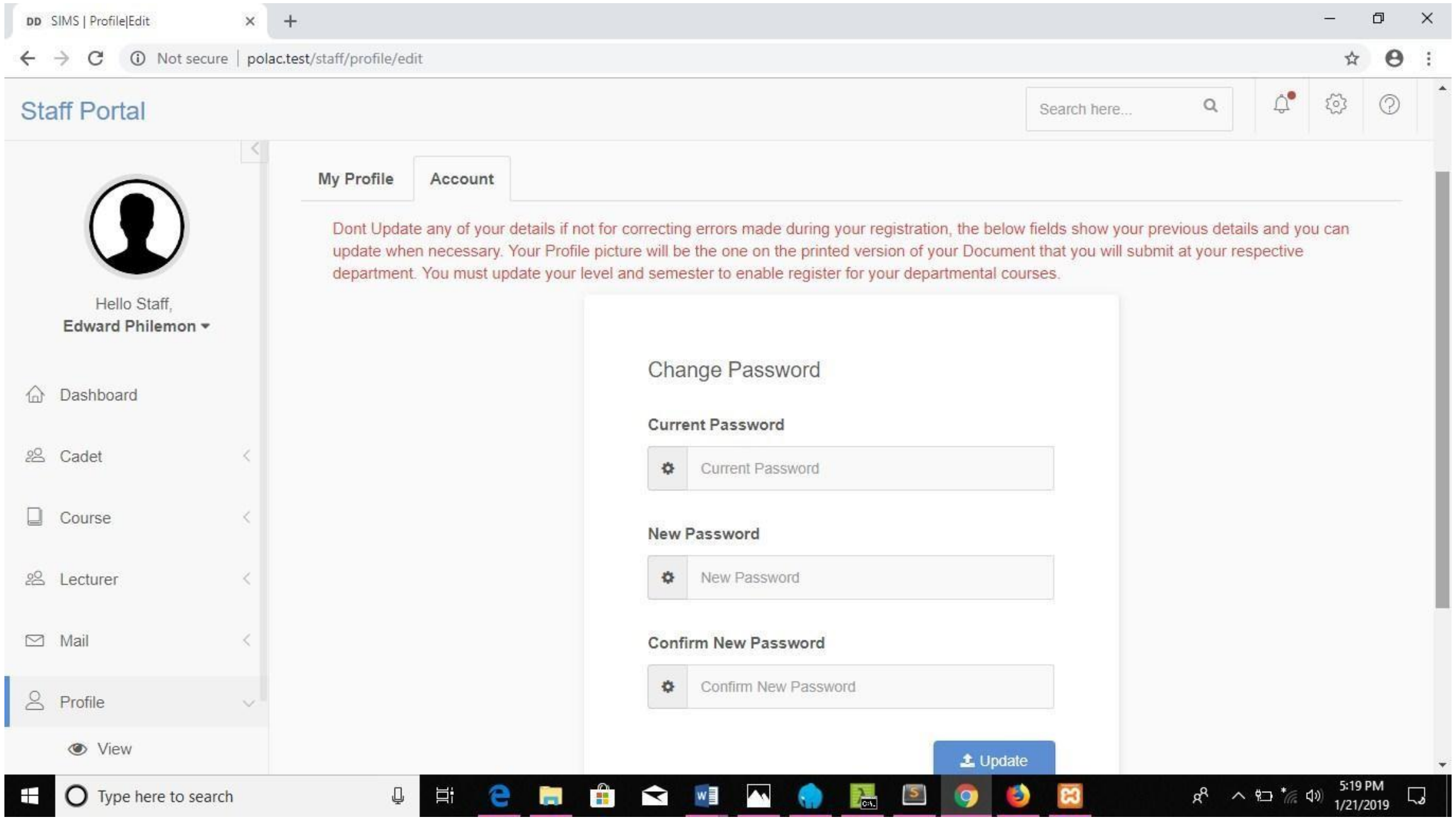

Figure 4. 20 Change password.

xviii. Lecturer dashboard

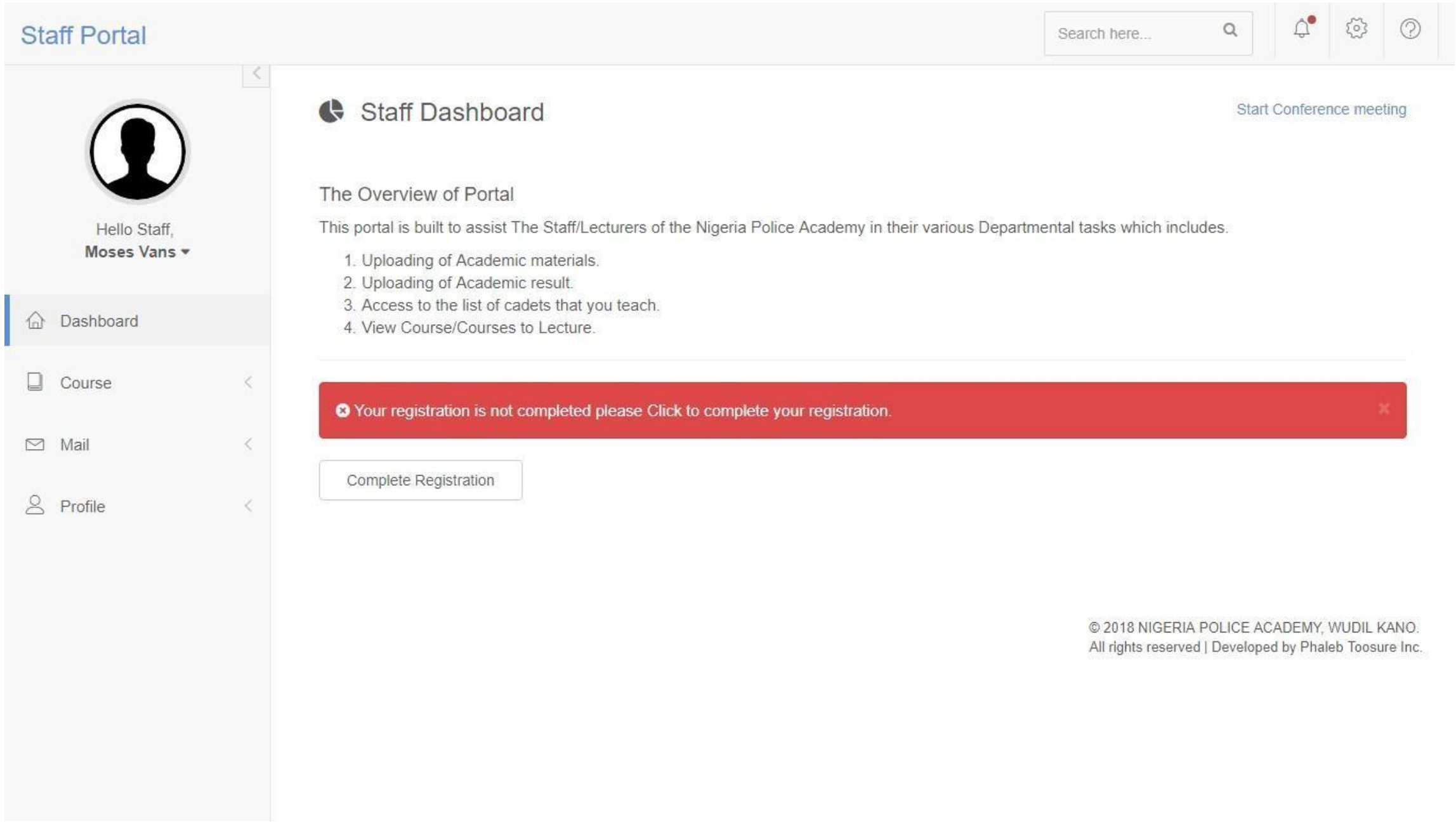

Figure 4. 21 Lecturer dashboard.

xix. Lecturer roles

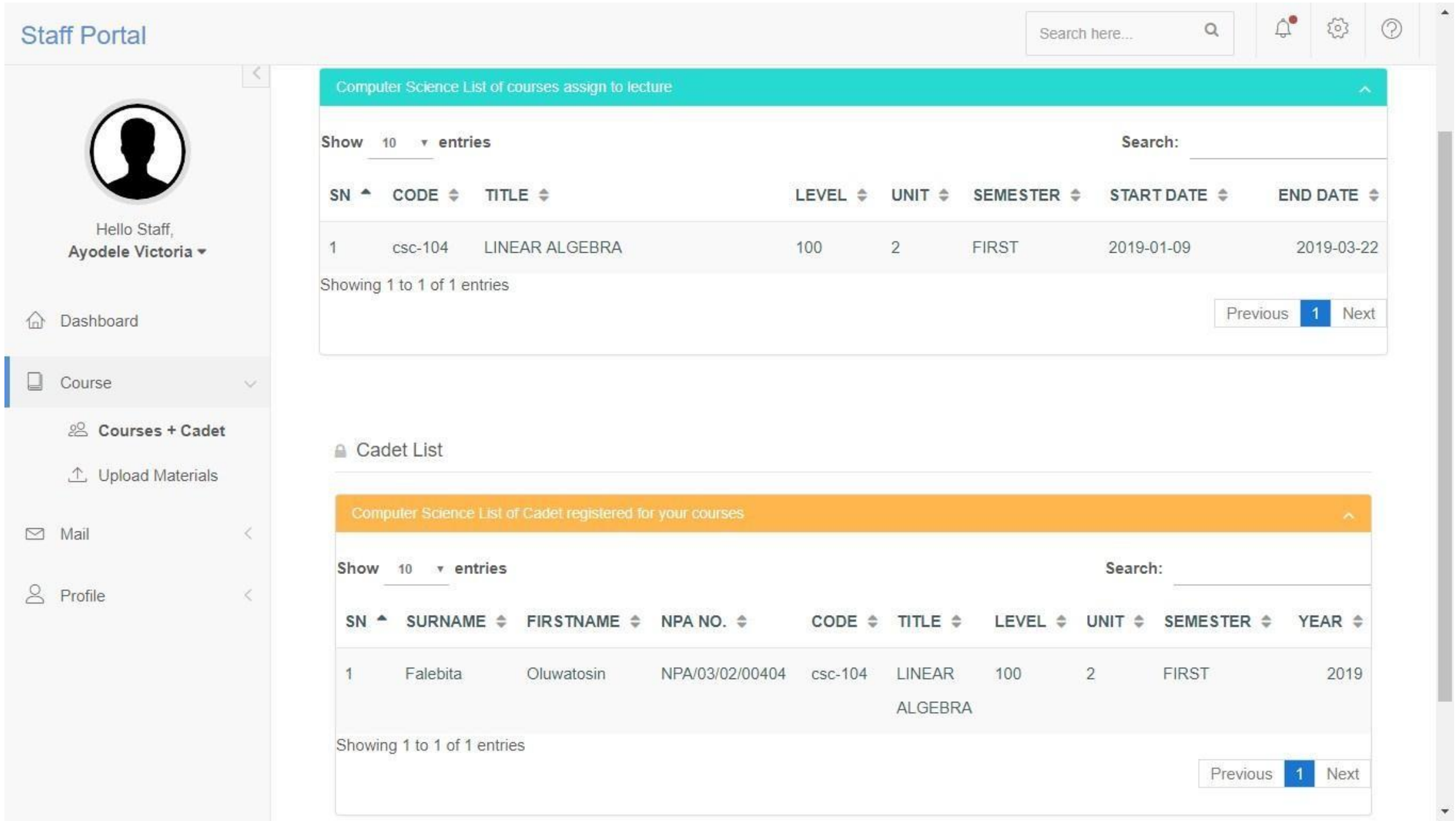

Figure 4. 22 List of Cadets registered for a course

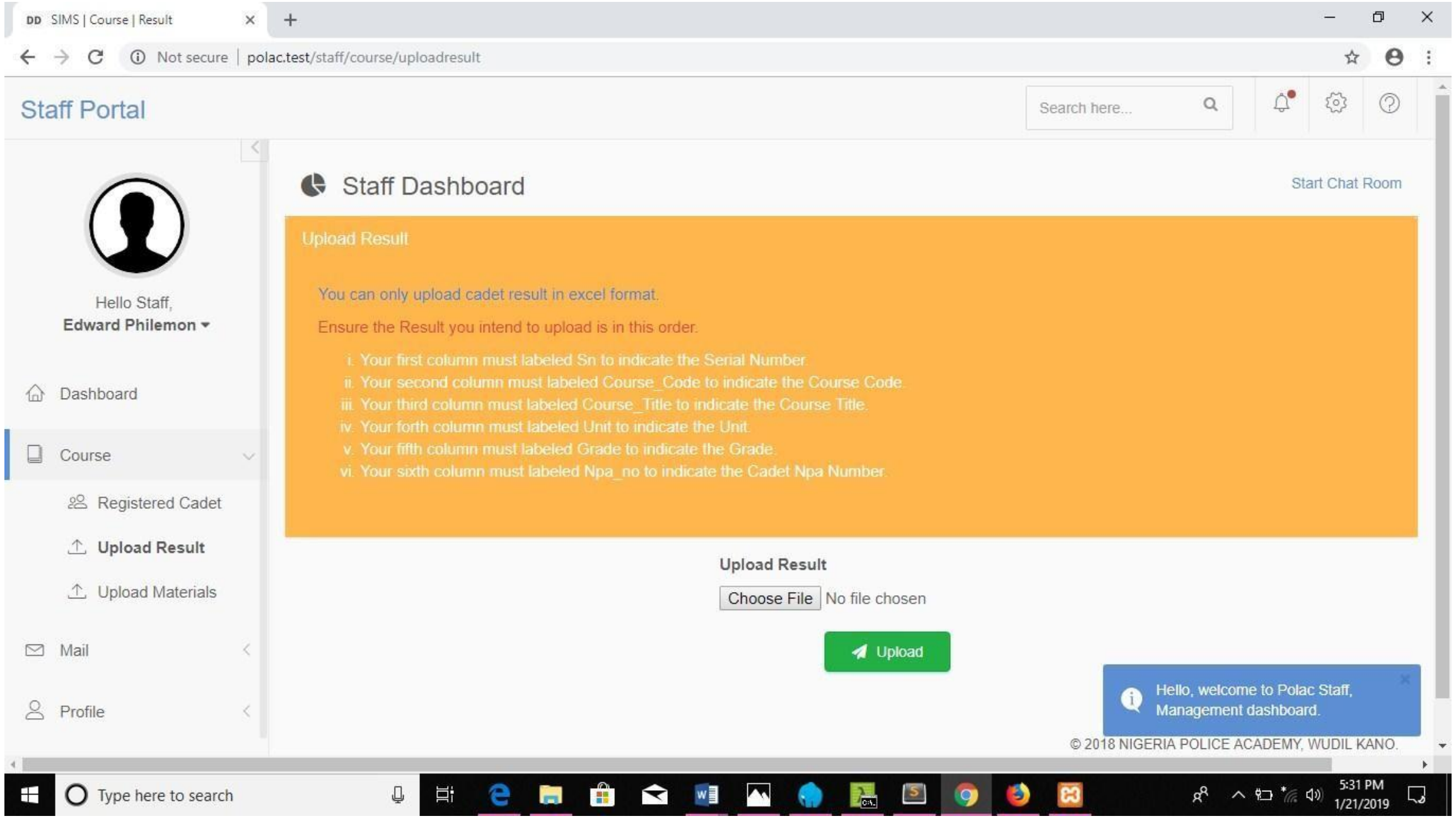

Figure 4. 23 Upload Cadet Scores.

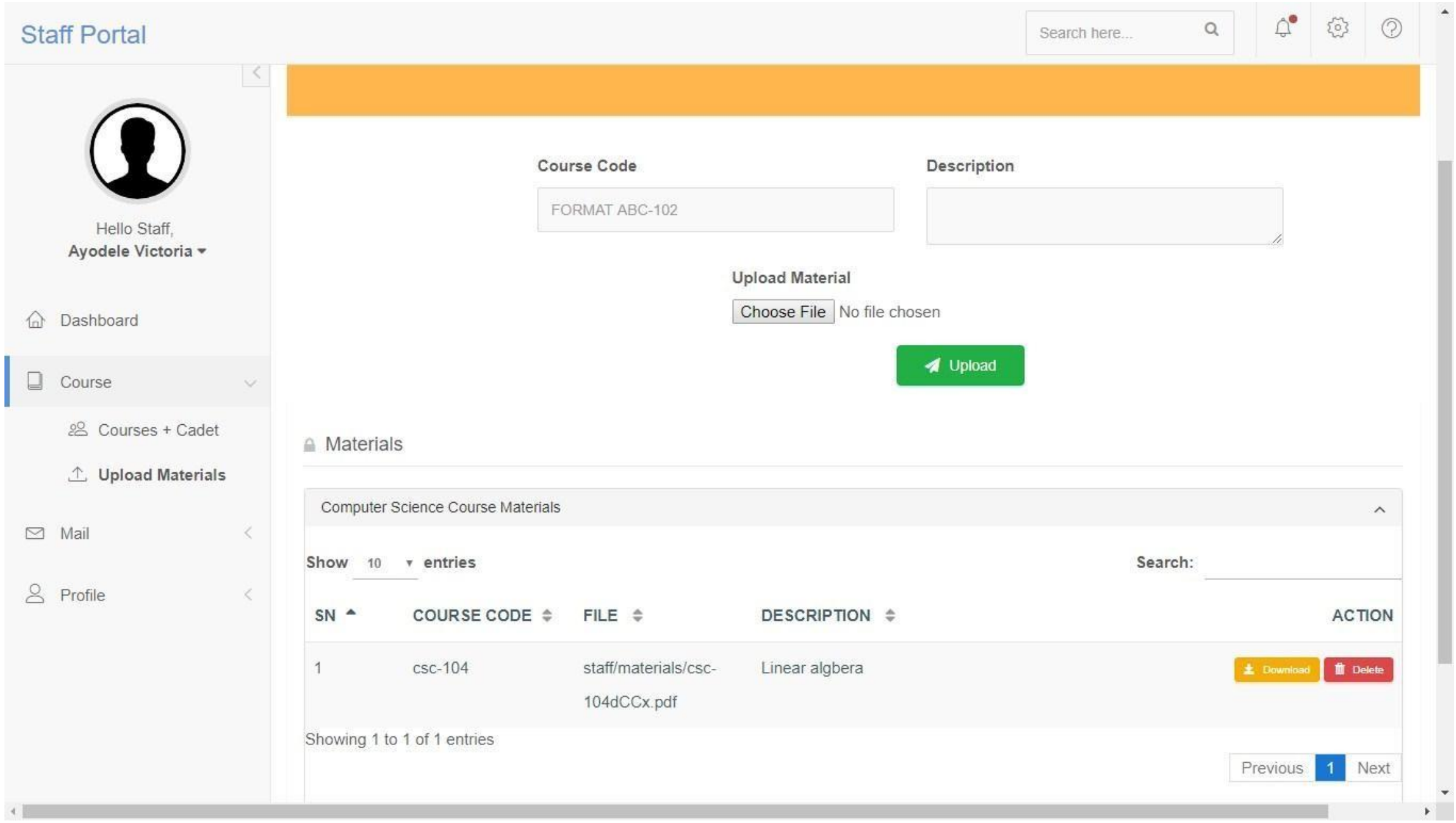

Figure 4. 24 Upload material.

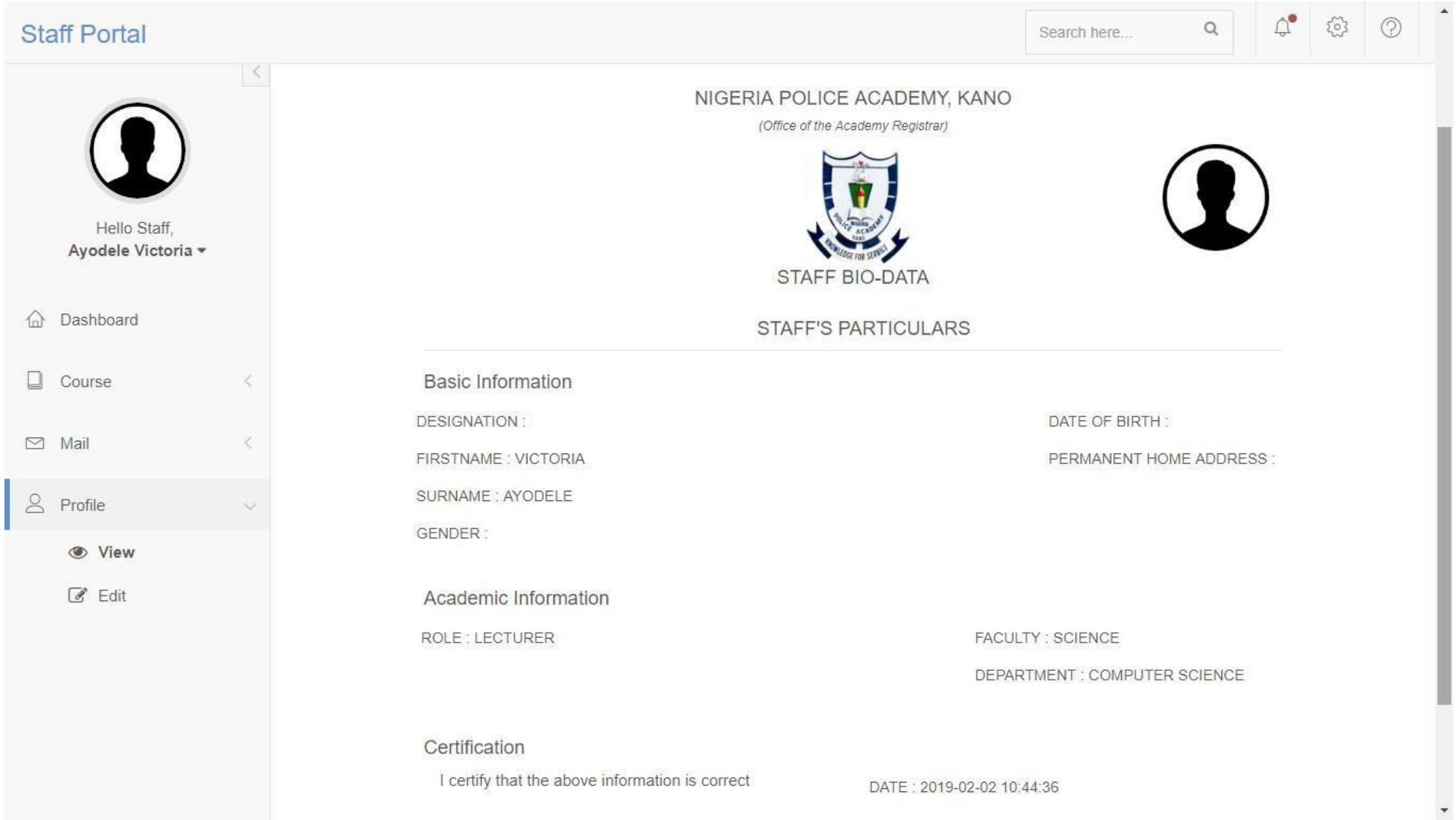

Figure 4. 25 Profile.

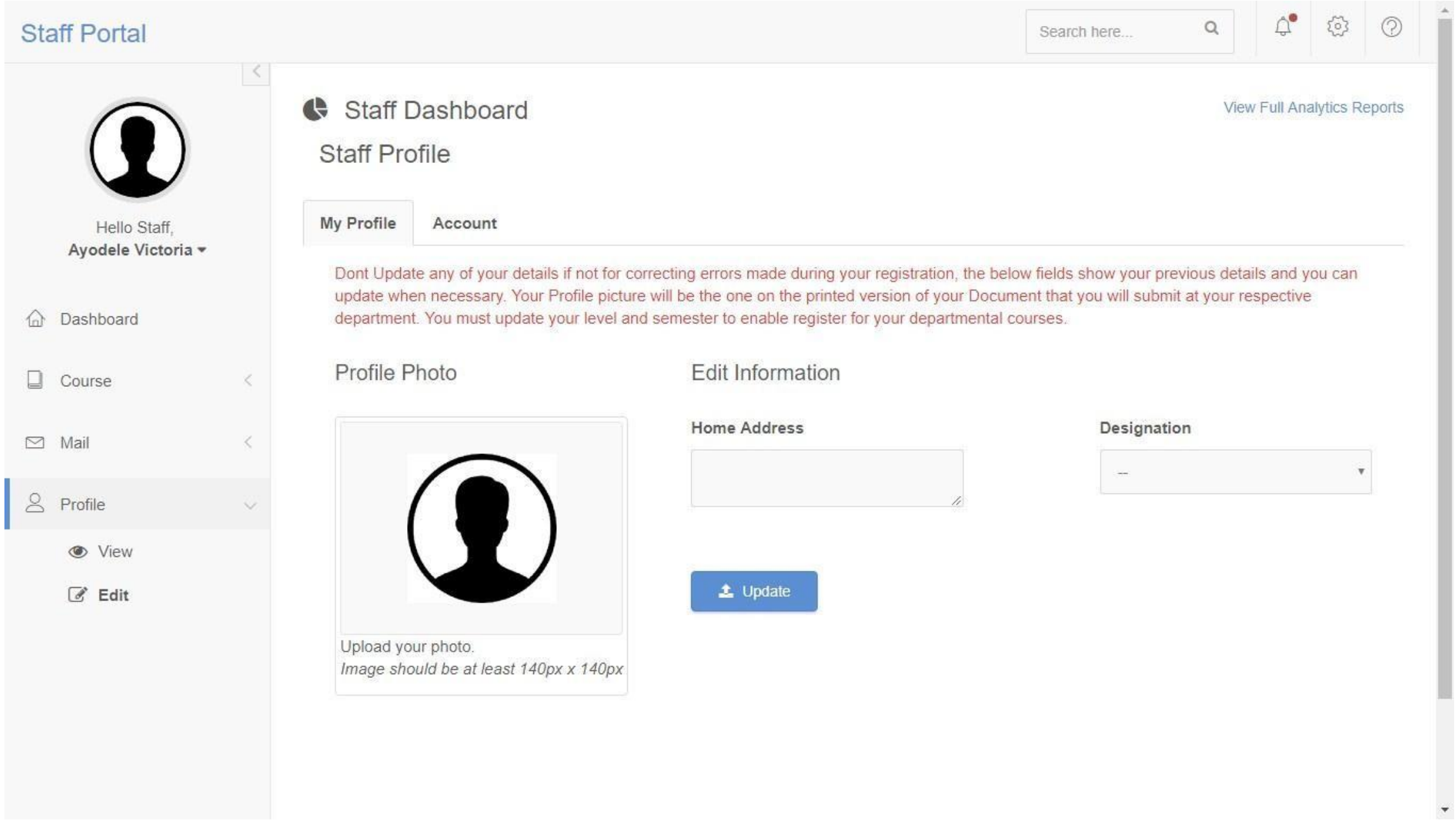

Figure 4. 26 Update profile.

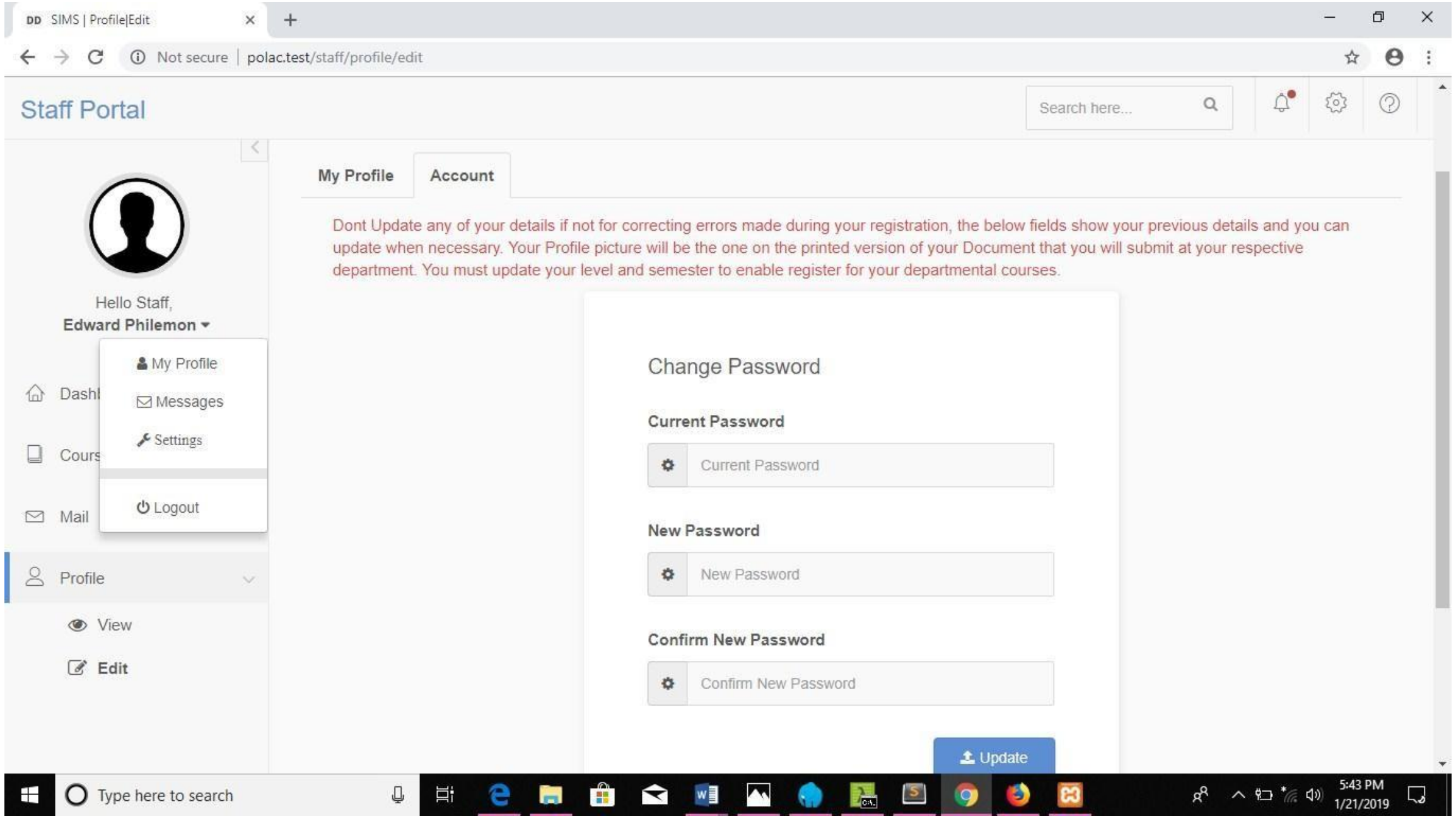

Figure 4. 27 Change password.

xx. Cadet Portal

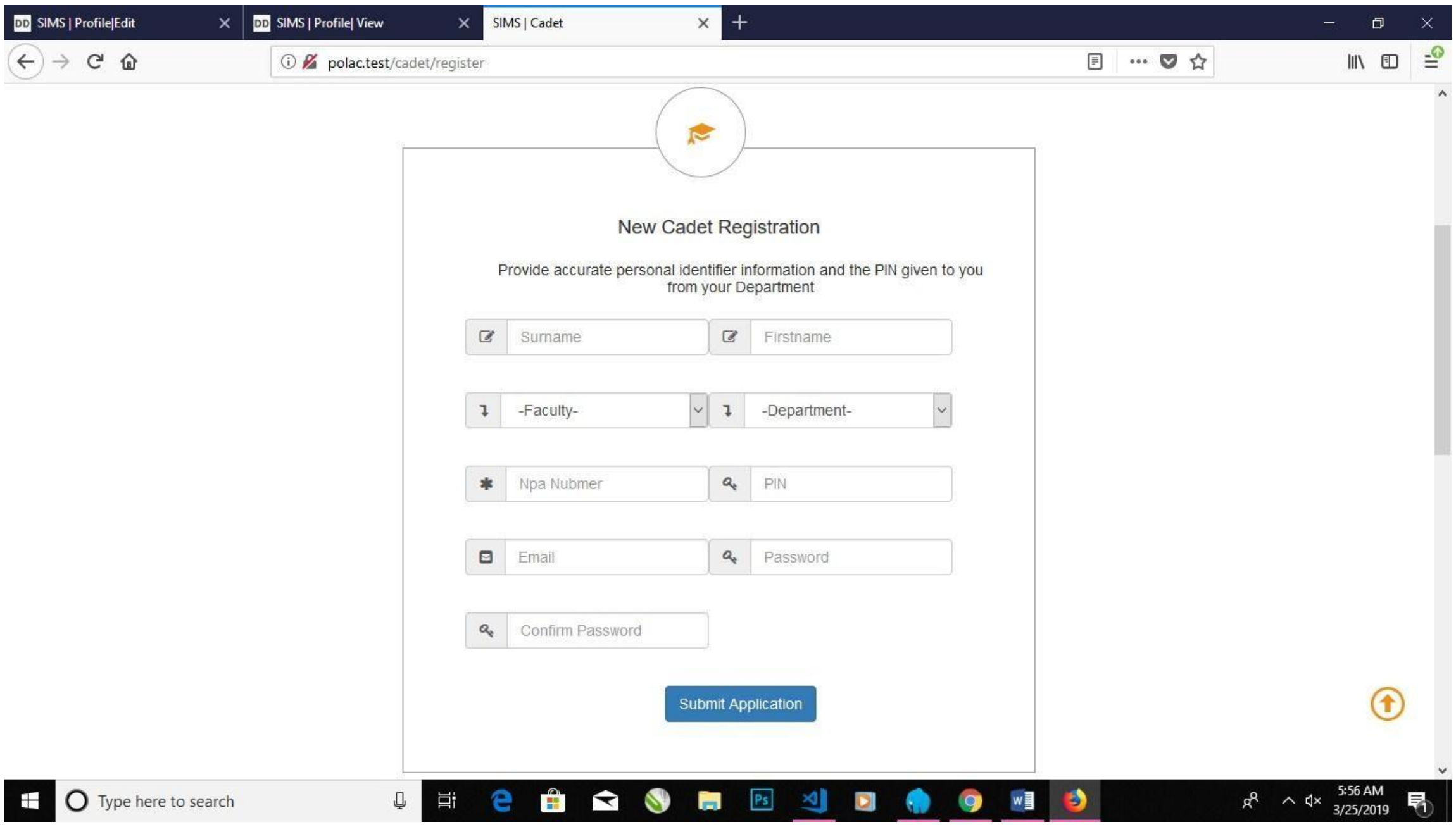

Figure 4. 28 Cadet registration.

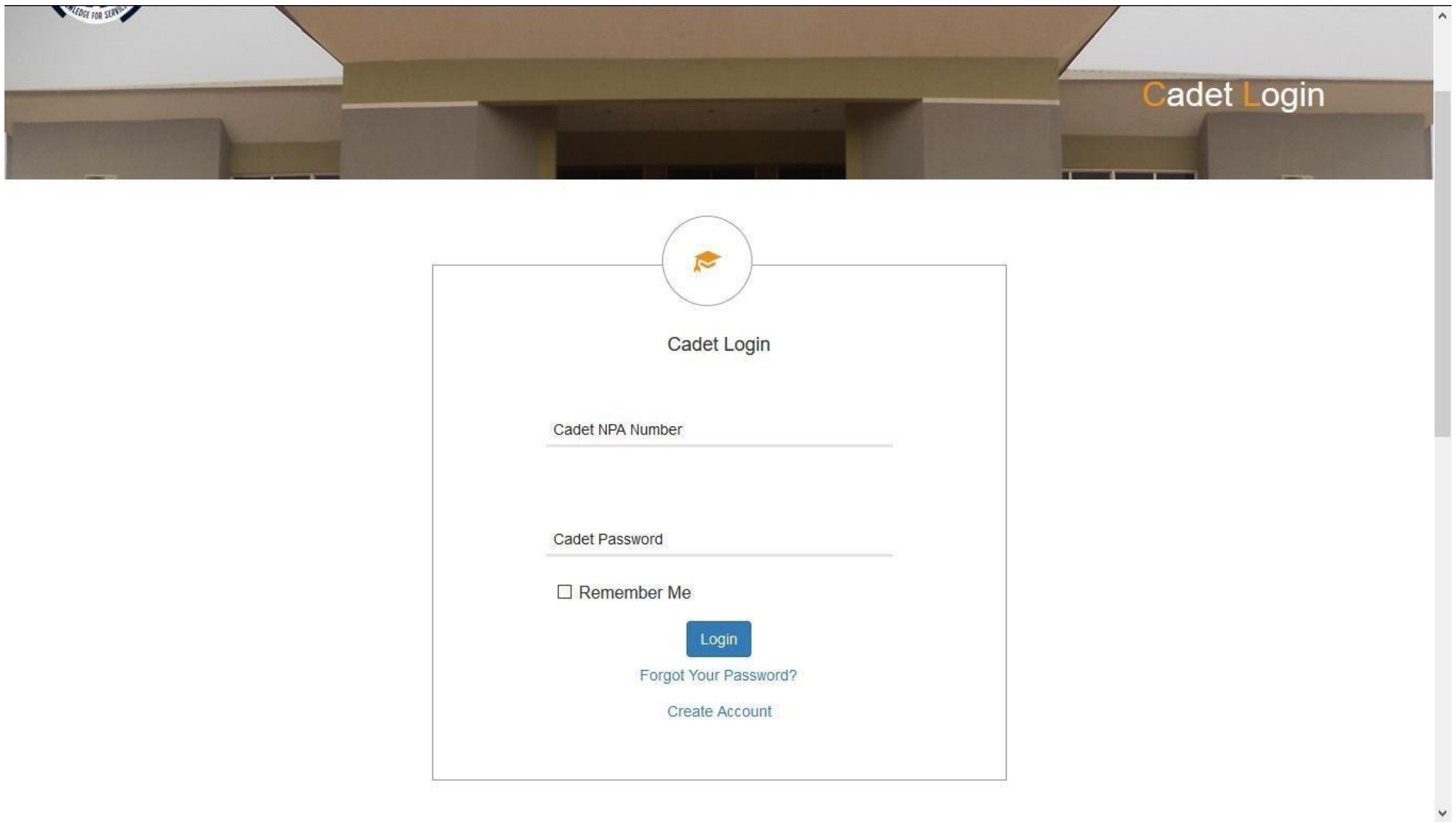

Figure 4. 29 Cadet login.

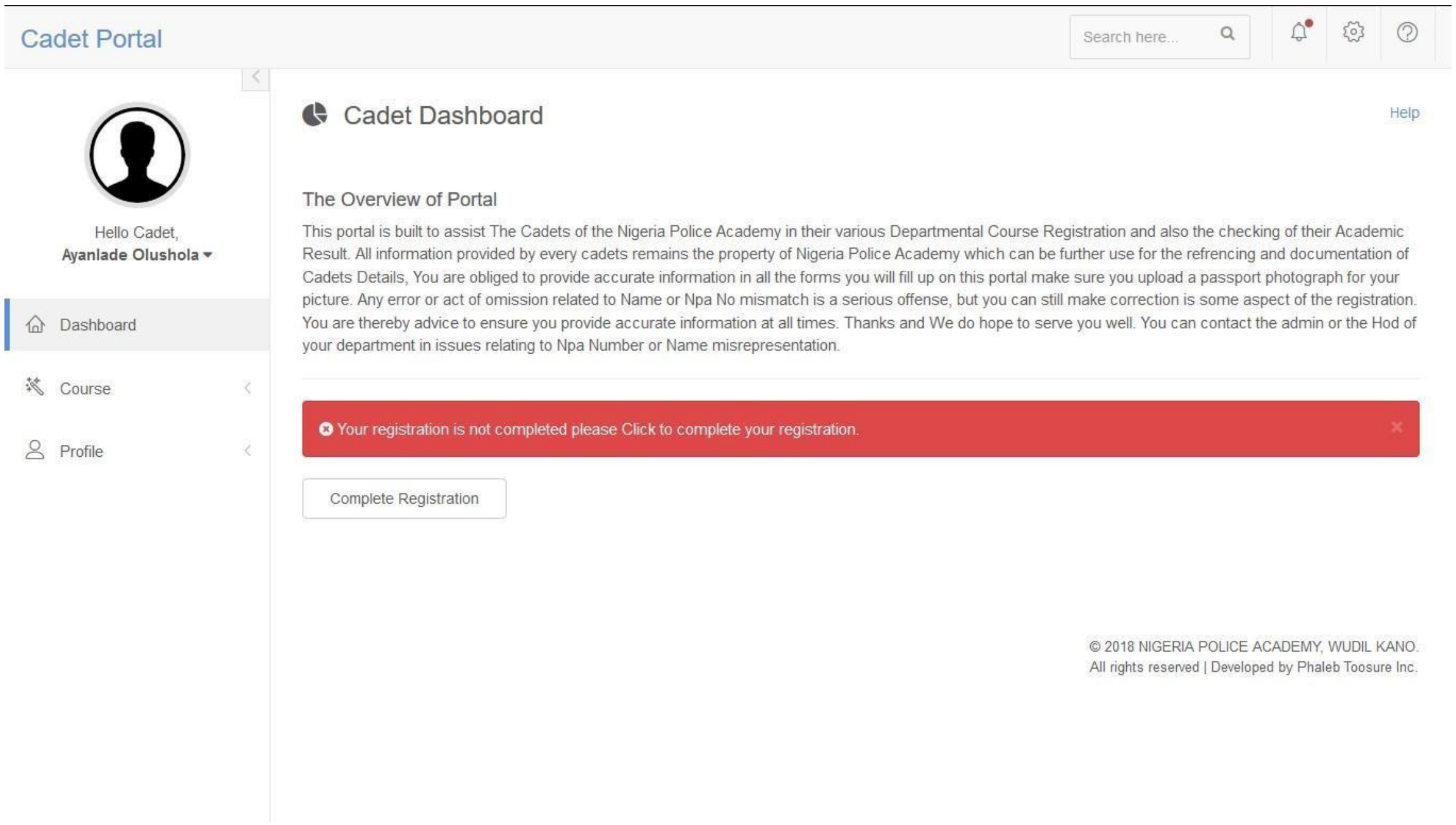

Figure 4. 30 Cadet dashboard.

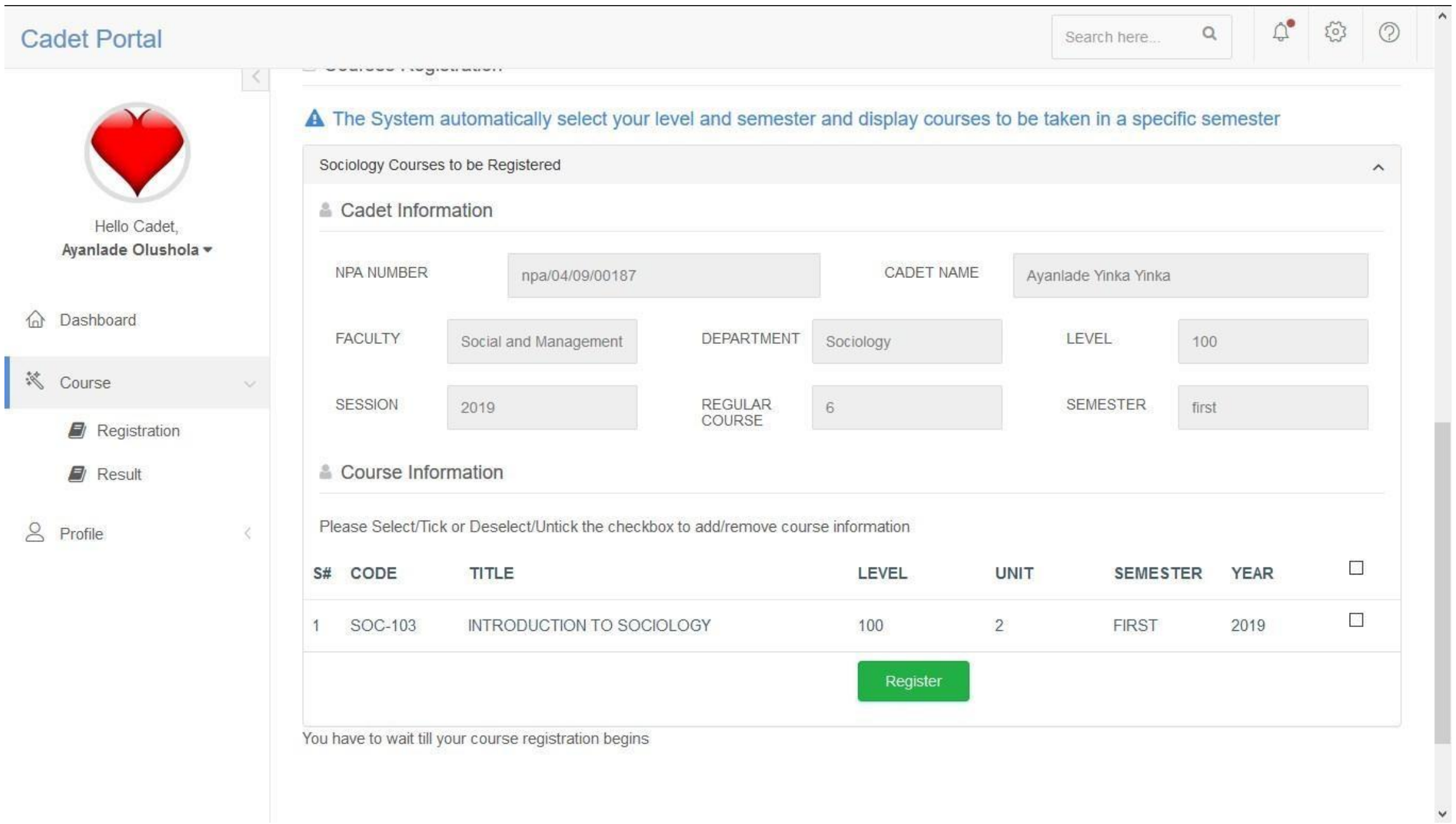

Figure 4. 31 Course registration..

# CHAPTER FIVE

# SUMMARY, CONCLUSION AND RECOMMENDATIONS

## 5.1 Summary

Keeping Information remains the mainstay in every organization, however the security of such information depends on the level of organization ranging from government to private organization. This paper assists in the automation of the existing paper-based information system for managing student and staff information in Nigeria Police Academy. The security of the web-based system was considered at the developmental stage of the application, all laravel security mechanism were implemented with access control level to maintain access to each portal of the application. This application provide a secure web based system to manage all information related to academic staff and students.

## 5.2 Conclusion

Information management has been a major concern to organization, different methods were used to keep information such as physical storage and file storage approach, the existing paper-based information system used in managing information in Nigeria Police Academy is obsolete and the academy suffers a large draw back in maintaining students record, access to relevant student information leads to a cumbersome task. This proposed system was able to provide a reliable and efficient approach for student and staff details, moreover the system was able to provide student with a web based course registration alongside course material for each course, results, breakdown of the maintenance courses for their academic degree program. To sum it up tasks performed by academic staff. i.e. (Hod & Lecturers) were fully automated such as assigning of course to lecturers and uploading of results e.t.c.

This paper designs and implements a secure web-based student information management system based on Laravel model, which achieved automated processing for all aspect of the existing paper-based information system. The experimental and simulation proved, web design based on Laravel framework, has scalability and robust scalability and ensure security, so as to improve the existing system.

## 5.3 Recommendation

The following are the recommendation required in order to effectively enhance the new system

i. Frequent Upgrade of the application is required in other to ensure security against latest web application vulnerabilities since new threats surface daily.

The following recommendations are for further study.

i. Provide an API(Application Programming Interface) that can be used by other developers to communicate with the website i.e. using Ionic-cordova to build an Andriod/IOS APP that will be compatible with the system.
ii. Automate the paper-based information system used in the Police wing of the academy.

# Appendix

## Source Code

Cadet.php : The code below shows the source code for the cadet model.

```
<?php

namespace App;

use App\Notifications\CadetResetPassword;
use Illuminate\Notifications\Notifiable;
use Illuminate\Foundation\Auth\User as Authenticatable;

class Cadet extends Authenticatable
{
    use Notifiable;

    /**
     * The attributes that are mass assignable.
     *
     * @var array
     */
     /**
      *  Changing of default database to pwrite
      */


    protected $fillable = [
        'id',
        'surName',
        'firstName',
        'middleName',
        'npaNumber',
        'pin',
        'email',
        'rc',
        'faculty',
        'department',
        'level',
        'semester',
        'squad',
```

```
    'sex',
    'dob',
    'homeTown',
    'localGovt',
    'state',
    'address',
    'nextOfKinSurName',
    'nextOfKinFirstName',
    'nextOfKinRelationship',
    'nextOfKinAddress',
    'passport',
    'password',

];

protected $hidden = [
    'password', 'remember_token',
];

/**
 * Send the password reset notification.
 *
 * @param  string  $token
 * @return void
 */
public function sendPasswordResetNotification($token)
{
    $this->notify(new CadetResetPassword($token));
}

/**
*
* One to many relationship between cadets
* and department
*/
public function depart()
{
    return $this->belongsTo('App\Depart', 'department','name');
}

/**
*
* many to many relationship between cadets
```

```
    * and course
    */

    public function regcourse()
    {
        return $this->belongsToMany('App\RegCourse', 'cadet_regcourses', 'cadet_id', 'regcourse_id')->withTimestamps();
    }

}
```

Faculty.php : The code below shows the faculty model.

```
<?php

namespace App;

use Illuminate\Database\Eloquent\Model;
use Illuminate\Database\Eloquent\SoftDeletes;

class Faculty extends Model
{

        use SoftDeletes;
    //

            public $incrementing = 'false';
            protected $keyType = 'string';
            protected $primaryKey = 'name';
    protected $fillable = [

                    'name',
    ];

    public function department()
    {
            return $this->hasMany('App\Department', 'faculty_name', 'name');
    }
}
```